\documentclass[
aps,prd,
nofootinbib,
superscriptaddress,
tightenlines,
amsmath,
amssymb
]{revtex4}

\usepackage{bm}
\usepackage{slashed}
\usepackage{float}
\usepackage[normalem]{ulem}

\usepackage{setspace}   

\usepackage{color, graphicx}   
\usepackage{float}      
\usepackage{rotating}   
\usepackage{caption}    
\usepackage{subcaption} 
\usepackage{amsmath}    
\usepackage{amssymb}    
\usepackage{mathtools}  
\usepackage{hyperref}   
\usepackage{cancel}     
\usepackage{listings}   
\usepackage{array}      
\usepackage{wrapfig}    
\usepackage{comment}    
\newenvironment{Figure}
{\par\medskip\noindent\minipage{\linewidth}}
{\endminipage\par\medskip}

\newcommand{\myprod}[2]{(#1\,#2)}

\newcounter{comment}

\newcommand{\fixedfigure}[3]{
    \begin{Figure}
        \centering
        \captionsetup{type=figure}
        \includegraphics[width=.5\linewidth]{#1}
        \captionof{figure}{#2}
        \label{#3}
    \end{Figure}
}

\allowdisplaybreaks
\begin{document}
\title{Likelihood \& Correlation
Analysis of Compton Form Factors for 
Deeply Virtual Exclusive Scattering on the Nucleon}

\author{Douglas Q. Adams} 
\email{yax6jr@virginia.edu}
\affiliation{Department of Physics, University of Virginia, Charlottesville, VA 22904, USA.}

\author{Joshua Bautista} 
\affiliation{Department of Physics, University of Virginia, Charlottesville, VA 22904, USA.}

\author{Marija \v Cui\'c} 
\affiliation{Department of Physics, University of Virginia, Charlottesville, VA 22904, USA.}

\author{Adil Khawaja} 
\affiliation{Department of Physics, University of Virginia, Charlottesville, VA 22904, USA.}

\author{Saraswati Pandey} 
\affiliation{Department of Physics, University of Virginia, Charlottesville, VA 22904, USA.}

\author{Zaki Panjsheeri}
\affiliation{Department of Physics, University of Virginia, Charlottesville, VA 22904, USA.}

\author{Gia-Wei Chern} 
\email{gc6u@virginia.edu}
\affiliation{Department of Physics, University of Virginia, Charlottesville, VA 22904, USA.}

\author{Yaohang Li} 
\affiliation{Old Dominion University}

\author{Simonetta Liuti} 
\email{sl4y@virginia.edu}
\affiliation{Department of Physics, University of Virginia, Charlottesville, VA 22904, USA.}

\author{Marie Bo\"{e}r} 
\affiliation{Virginia Tech}

\author{Michael Engelhardt} 
\affiliation{Department of Physics, New Mexico State University, Las Cruces, NM 88003, USA}

\author{Gary R. Goldstein} 
\affiliation{Tufts University}

\author{Huey-Wen Lin} 
\affiliation{Michigan State University}

\author{Matthew D. Sievert} 
\affiliation{Department of Physics, New Mexico State University, Las Cruces, NM 88003, USA}

\vspace{0.5cm}
\collaboration{EXCLAIM Collaboration}

\vspace{0.5cm}
\begin{abstract}    
A likelihood analysis of the 
observables in deeply virtual exclusive photoproduction off a proton target, $ep \rightarrow e' p' \gamma'$, is presented. 
Two processes contribute to the reaction: deeply virtual Compton scattering, where the photon is produced at the proton vertex, and the Bether-Heitler process, where the photon is radiated from the electron.
We consider the unpolarized process for which the largest amount of data with all the kinematic dependences are available from 
corresponding datasets with unpolarized beams and unpolarized targets from Jefferson Lab.
We provide and use a method which derives a joint likelihood of the Compton form factors, which parametrize the deeply virtual Compton scattering amplitude in QCD, for each observed combination of the kinematic variables defining the reaction. 
The unpolarized twist-two cross section likelihood fully constrains only three of the Compton form factors (CFFs).
The impact of the twist-three corrections to the analysis is also explored.
The derived likelihoods are explored using Markov chain Monte Carlo (MCMC) methods.
Using our proposed method we derive CFF error bars and covariances. 
Additionally, we explore methods which may reduce the magnitude of error bars/contours in the future.
\end{abstract}

\maketitle

\section{Introduction}
\label{sec:intro}
The process of deeply virtual Compton scattering (DVCS), where an electron scatters coherently off a proton target  producing a hard photon, 
is deemed the most accurate probe of the nucleon quark and gluon distributions in coordinate space.  The latter can be extracted from the non-forward QCD matrix elements between the initial ($p$) and final ($p'$) proton, known as generalized parton distributions (GPDs), through Fourier transformation with respect to the proton momentum transfer ($\Delta=p-p'$). Assuming the validity of QCD factorization, one can parametrize the DVCS amplitude in terms of several Compton form factors (CFFs), which arise from the various allowed beam-target polarization configurations. CFFs are convolution integrals of GPDs with known kernels/coefficient functions. GPDs can therefore be inferred, although only indirectly,  from the experimental observation of CFFs (we refer the reader to reviews on the subject in \cite{Diehl:2001pm,Belitsky:2005qn,Kumericki:2016ehc} and references therein). 

DVCS holds a unique role among all deeply virtual exclusive scattering (DVES) experiments since, through the interference with the background Bethe Heitler (BH)
process, where the final state hard photon is radiated from the electron, one can extract the CFFs directly at the amplitude level.  
Furthermore, because of the presence of only one distinct hadronic blob, deeply virtual electron photoproduction is a clean probe, as compared to deeply virtual exclusive meson production, or to similar processes with more than one hadron in the final state, where an additional non leading dependence on the QCD scale, $Q^2$, is more likely to contaminate a QCD-based analysis.
Notwithstanding, a quantitative determination of CFFs, and consequently of GPDs, from DVCS data has been notoriously difficult, and no definitive result exists to date on the dominance of the asymptotic regime that should be manifested at $t/Q^2 \rightarrow 0$, with $t=\Delta^2$.  
A major hurdle in the analysis to extract GPDs from experiment arises due to the presence of two inverse problems: a first one for the extraction of the various CFFs from data, and a subsequent one concerning the extraction of GPDs from CFFs.

In this paper we present results of a joint multivariate Bayesian likelihood analysis of unpolarized electron scattering off an unpolarized proton target using a subset of the available DVCS data which displays a full kinematic dependence in all variables, {\it i.e.} which is not integrated over the azimuthal angle, $\phi$. Our aim is to place quantitative bounds on the experimental extraction of CFFs. We define the unknown CFFs as parameters and we first assume that their kinematic coefficients, as well as the proton elastic form factors from the BH amplitude are exactly calculable in a QCD framework. We have used all unpolarized DVCS data for our analysis which consequently displays full covariance in our results. In this respect, we are adopting what could be considered as a ``brute force", or completely unbiased method. 
The first and foremost goal of this analysis is to answer the questions: are present data consistent with QCD? 
Is there enough evidence in the data to validate a QCD-based description of deeply virtual exclusive processes? Are present experimental data following trends expected in the asymptotic regime, or are higher twist terms dominant? What further methods would be appropriate for testing the QCD nature of the cross section?

In the $t/Q^2 \rightarrow 0$ limit, the QCD evolution of GPDs is expected to proceed analogously as for the parton distribution functions (PDFs) and distribution amplitude (DAs) from inclusive deep inelastic scattering experiments,  {\it i.e.} in a collinear framework, generating $Q^2$ dependent logarithmic corrections. There is only a technical point of departure from PDFs and DAs given by the fact that the kernels of the evolution equations, as well as the next-to-leading order Wilson coefficient functions, also depend on the GPDs' skewness variable, $\xi$, measuring the longitudinal momentum fraction difference between the initial and final struck particles, besides the standard longitudinal parton momentum fraction $x$. 
Departures from the parton model-based description, parametrized in QCD by twist-two GPDs, could originate kinematically as target mass corrections, as well as dynamically, due to the exchange of gluons between the struck quark and
the spectator system, and to coherent phenomena possibly including multi-quark scattering, resonance production, and scattering from constituent mesons. All of these phenomena are potential higher twist effects displaying characteristic $Q^2$ inverse powers dependence. 

By illustrating several issues that arise for the specific dataset analyzed here, we are seeking to establish a statistical framework and a general procedure that can be subsequently applied to all DVES-type data.  
The paper is organized as follows: in Section \ref{sec:2} we give a theoretical description of the cross section using a QCD-based formulation up to twist-three terms, as a point of reference. We introduce the main motivation of our analysis, which is that the present DVCS cross section data allows equivalently good fits with completely different CFFs.  
We subsequently introduce a systematic approach that allows us to address the several open questions stemming from the result of using a simple fitting procedure, including: how many form factors can be extracted from data, and what are their boundaries? How can the given theoretical background be validated? The quantitative  methods to address these questions require a full-fledged multivariate likelihood analysis and they are described in Section \ref{sec:3}; in Section \ref{sec:4} we present and discuss the numerical results of this analysis, addressing the question of whether the data are \textcolor{blue}{well described by current cross section models, as well as} interpretable in a QCD context including the contributions of higher twists. 
The conclusions and outlook are given in Section \ref{sec:conclusions}.


\section{Theoretical Framework and Motivation}
\label{sec:2}
\begin{figure}
    \centering
    \includegraphics[width=10cm]{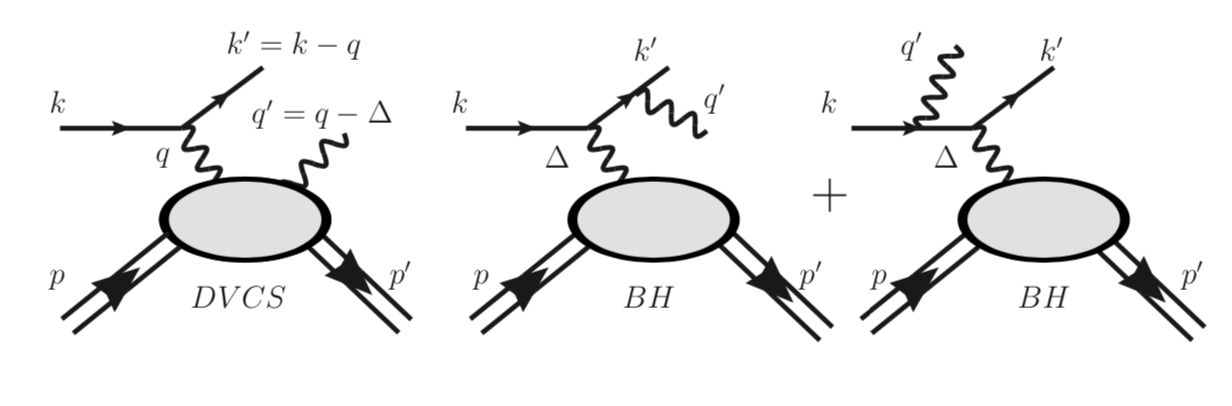}
    \caption{Leading order diagrams for exclusive photon electro-production $ep \rightarrow ep\gamma$. The DVCS process is parametrized by GPDs, while the Bethe-Heitler background process is parametrized by elastic form factors.}
    \label{fig:feynman-diagram}
\end{figure}

The exclusive photon electroproduction process $ep \rightarrow e' p' \gamma'$ can occur either through deeply virtual Compton scattering (DVCS) or through the Bethe-Heitler (BH) process (see Fig.~\ref{fig:feynman-diagram}). 
The latter are summed coherently at the amplitude level, giving rise to the following differential cross section
\begin{equation}
\label{eq:total-cross-section1}
    \frac{d^4 \sigma}{dx_{Bj} dQ^2 dt d\phi} \equiv \sigma_{TOT} = \sigma_{BH} + \sigma_{DVCS} + \sigma_{\mathcal{I}},
\end{equation}
where $\sigma_{BH}$ and $\sigma_{DVCS}$ are the Bethe-Heitler and DVCS contributions, respectively, while $\sigma_{\mathcal{I}}$ is the cross section resulting from their interference. 
In Eq.\eqref{eq:total-cross-section1}, the cross section is written in terms the kinematic variables $(s, Q^2, x_{Bj}, t, \phi)$, where
\begin{itemize}
    \item $s=(k+p)^2$ is the electron-proton center of mass energy squared,
    \item $Q^2 = -(k-k')^2 = - q^2$ is the four-momentum transfer squared between the incoming and outgoing electrons,
    \item $x_{Bj} = Q^2 / 2(pq)$ is Bjorken-$x$. In the asymptotic limit, disregarding $t/Q^2$ and $M^2/Q^2$ corrections, $x_{Bj}$ is written in terms the skewness parameter, $\xi = - (\Delta q)/[(pq)+(p'q)]= x_{Bj}/(2- x_{Bj})$
    \item $t = (p- p')^2 = (q'-q)^2$ is the four-momentum transfer squared between the initial and final protons ($q'$ is the final photon four-momentum),
    \item $\phi$ is the azimuthal angle between the planes defined by the electron momenta, by the final proton, $p'$, and photon, $q'$.
\end{itemize}
In this paper we will focus on unpolarized electrons scattering off an unpolarized proton, therefore the cross section does not depend on an additional azimuthal angle stemming from the proton spin orientation. 
(Note that \cite{Kriesten:2019jep} writes the five-fold differential cross section, which includes the transverse spin angle $\phi_S$, while Eq.~\eqref{eq:total-cross-section1} is integrated over $\phi_S$.
Therefore, the $\Gamma$ factor used in the expressions following Eq.~\eqref{eq:total-cross-section1} has an extra factor of $2\pi$ compared to the $\Gamma$ factor in \cite{Kriesten:2019jep, Kriesten:2020apm, Kriesten:2020wcx}. 
In this case, the BH contribution to the cross section is parametrized in terms of the proton Dirac and Pauli elastic form factors $F_1(t)$ and $F_2(t)$, as follows \cite{Kriesten:2019jep},
\begin{equation}
\label{eq:sigmaBH1}
    \sigma_{BH}=\frac{\Gamma}{t} \, \bigg\{ A_{BH} \left[F_1^2(t)+\tau F_2^2(t) \right]+B_{BH} \,  \tau \left[F_1(t)+F_2(t)\right]^2  \bigg\} \, ,
\end{equation}
where $\tau = -t/4M^2$, $M$ being the proton mass. The pre-factor, $\Gamma$, includes the Jacobian for the change of variables in going from the final state particles' momenta to the invariants defined above, and it is expressed as \cite{Kriesten:2019jep},
\begin{equation}
\label{eq:Gamma}
\Gamma = \frac{\alpha^3 (2 \pi) }{16\pi^2 (s-M^2)^2 \sqrt{1+\gamma^2}\, x_{Bj} },
\end{equation}
where $\alpha$ is the electromagnetic fine structure constant, and $\gamma^2 = 4M^2x_{Bj}^2/Q^2$. 
The coefficients $A_{BH}, B_{BH}$ are dimensionless functions of the kinematic variables $(s, Q^2, x_{Bj}, t, \phi)$ that were calculated in \cite{Kriesten:2019jep} (we present them for completeness  in Appendix \ref{app:A}).
The nucleon elastic form factors are known precisely in a wide kinematic range (for more details, see the review \cite{Punjabi:2015bba} an references therein), and various parametrizations for them exist in the literature \cite{Kelly:2004hm, Bradford:2006yz, Brash:2001qq}.
The BH contribution is therefore considered a known term that we will subtract out from the data in our analysis.

Assuming a QCD factorized framework, the DVCS contribution is described by four complex Compton form factors (CFFs) at leading order: $\mathcal{H}$, $\mathcal{E}$, $\widetilde{\mathcal{H}}$ and $\widetilde{\mathcal{E}}$, which correspond to the various allowed quark-proton polarization configurations in the proton. 
Information regarding the 3D partonic structure of nucleons is encoded in the QCD correlation functions that define the generalized parton distributions (GPDs) \cite{Ji:1996nm,Radyushkin:1997ki,Muller:1994ses} (for comprehensive reviews of GPDs and DVCS we refer the reader to \cite{Diehl:2003ny, Belitsky:2005qn, Kumericki:2016ehc}).
In a nutshell, GPDs depend on the longitudinal momentum fraction, $x$, on the momentum transfer, $t$, between the initial and final proton -- allowing us to access the quark and gluon transverse spatial distribution through Fourier transformation\cite{Diehl:2002he, Burkardt:2002hr, Soper:1976jc} -- and on the skewness parameter, $\xi$.
The second Mellin moments  of the GPDs in the variable $x$, are directly related to the form factors parametrizing the QCD energy momentum tensor, whose matrix elements between proton states describe orbital angular momentum, pressure, and the shear forces, that emerge from the proton's  quark and gluon internal dynamics  \cite{Ji:1996ek, Polyakov:2002yz,Shanahan:2018nnv,Lorce:2021xku}. 
GPDs enter the DVCS amplitude in convolutions with perturbative QCD Wilson coefficient functions, giving the CFFs,
\begin{equation}
\label{eq:CFFdefn1}
\begin{split}
    \mathcal{F}(\xi, t, Q^2) &= \sum_q e_q^2 \int_{-1}^{+1} d x\left[\frac{1}{\xi-x-i \epsilon}-\frac{1}{\xi+x-i \epsilon}\right] F^q(x, \xi, t, Q^2) \\
    \widetilde{\mathcal{F}}(\xi, t, Q^2) &= \sum_q e_q^2 \int_{-1}^{+1} d x\left[\frac{1}{\xi-x-i \epsilon}+\frac{1}{\xi+x-i \epsilon}\right] \widetilde{F}^q(x, \xi, t, Q^2)
\end{split}
\end{equation}
where $F^q = \{H^q, E^q\}$, $\widetilde{F}^q = \{\widetilde{H}^q, \widetilde{E}^q\}$, and $e_q$ is the fractional charge of the quark $q$ (analogous expressions can be written for gluons).

Using the covariant helicity amplitude formalism described in detail in \cite{Kriesten:2019jep} with the virtual photon, $q$, on the negative $z$-axis \cite{Diehl:2005pc}, the expression for the unpolarized cross section at leading twist in this formalism is written as (see also \cite{Kriesten:2020wcx, Kriesten:2020apm, Almaeen:2022imx, Almaeen:2024guo})
\begin{eqnarray}
\label{eq:sigmaDVCS1}
\sigma_{D V C S}  & = & \frac{\Gamma}{Q^2(1-\epsilon)} 
\Bigg\{ 2( 1 - \xi^2) \bigg( \left[{\Re e \mathcal{H}}(x_{Bj},t,Q^2)\right]^2 + \left[{\Im m \mathcal{H}}(x_{Bj},t,Q^2)\right]^2 + \big[{ \Re e \widetilde{\mathcal{H}}}(x_{Bj},t,Q^2) \big]^2 + \big[{ \Im m \widetilde{\mathcal{H}}}(x_{Bj},t,Q^2)\big]^2\bigg) \nonumber \\
& + & \frac{t_0-t}{2 M^2}\bigg( \left[{ \Re e \mathcal{E}}(x_{Bj},t,Q^2) \right]^2 + \left[{ \Im m \mathcal{E}}(x_{Bj},t,Q^2)\right]^2 + 
\big[ \xi \, { \Re e  \widetilde{\mathcal{E}}(}x_{Bj},t,Q^2)\big]^2 +  \big[\xi  \,{ \Im m  \widetilde{\mathcal{E}}}(x_{Bj},t,Q^2) \big]^2 \bigg)   \nonumber \\
&  - & 4 \xi^2\bigg({\bf \Re e \mathcal{H}}(x_{Bj},t,Q^2) \, { \Re e \mathcal{E}}(x_{Bj},t,Q^2) + { \Im m \mathcal{H}}(x_{Bj},t,Q^2) \, \Im m \mathcal{E}(x_{Bj},t,Q^2) 
+ { \Re e \widetilde{\mathcal{H}}}(x_{Bj},t,Q^2) \, { \Re e \widetilde{\mathcal{E}}}(x_{Bj},t,Q^2) \nonumber \\
&+& { \Im m \widetilde{\mathcal{H}}(}x_{Bj},t,Q^2) \, { \Im m \widetilde{\mathcal{E}}}(x_{Bj},t,Q^2) \bigg)  \Bigg\} ,
\end{eqnarray}
where 
\begin{equation}
\epsilon = \frac{1-y-\frac{1}{4} y^2 \gamma^2}{1-y+\frac{1}{2} y^2+\frac{1}{4} y^2 \gamma^2},
\end{equation}
with $y=(pq)/(pk)= Q^2/(x_{Bj} s)$, is the ratio of the longitudinal to transverse virtual photon polarization evaluated in the $M^2/Q^2 <<1 $ limit, and $t_0 = - 4 M^2\xi^2/(1-\xi)$. The following remarks are in order: 
\begin{enumerate}
    \item the leading order DVCS contribution is a quadratic expression of all four complex CFFs corresponding to eight unknowns (we call all eight terms ``CFFs", for brevity); 
    \item since the unpolarized case involves only helicity amplitudes for a transversely polarized virtual photon whose phases cancel when multiplied with their complex conjugates, the expression in Eq.\eqref{eq:sigmaDVCS1} is  independent of the azimuthal angle, $\phi$; 
    \item the cross section depends on $s$ only through the overall factor, $\Gamma/Q^2(1-\epsilon)$.
\end{enumerate}


\vspace{0.3cm}
The BH-DVCS interference contribution, $\sigma_{\cal I}$, is dependent on $\phi$ and reads,
\begin{eqnarray}
\label{eq:sigmaINT1}
    \sigma_{\mathcal{I}} & = & \frac{\Gamma}{Q^2}\bigg\{A_{\mathcal{I}} \, \left[ F_1(t)\,  {\bf \Re e\mathcal{H}}(x_{Bj},t,Q^2)+ \tau \, F_2(t)\, {\bf \Re e\mathcal{E}}(x_{Bj},t,Q^2)\right] 
    +  B_{\mathcal{I}} \, [F_1(t)+F_2(t)] \, {\bf \Re e }\left[ {\bf \mathcal{H}}(x_{Bj},t,Q^2) + {\bf \mathcal{E}}(x_{Bj},t,Q^2) \right] 
\nonumber \\ 
& + & C_{\mathcal{I}} [F_1(t)+F_2(t)] \, {\bf \Re e }\widetilde{\bf \mathcal{H}}(x_{Bj},t,Q^2) \bigg\}  .
\end{eqnarray}
%
The kinematic coefficients $A_{\mathcal{I}}$, $B_{\mathcal{I}}$, $C_{\mathcal{I}}$, in Eq.\eqref{eq:sigmaINT1} are dimensionless functions of the kinematic variables $(s, Q^2, x_{Bj}, t, \phi)$; they have been calculated in \cite{Kriesten:2019jep, Kriesten:2020apm, Kriesten:2020wcx}; for completeness, we present their expressions  in Appendix \ref{app:B}. 
\textcolor{blue}{Their dependence on $\phi$ is discussed in Section \ref{sec:4}. }
\begin{enumerate}
    \item the interference term is parametrized by only three unknowns given by the real parts of the CFFs: $\Re e \mathcal{H}$, $\Re e \mathcal{E}$, and $\Re e \widetilde{\mathcal{H}}$, which enter the equation linearly in a form which closely tracks the elastic cross section parametrization into form factors 
    \cite{Sofiatti:2011yi};
    \item The $\phi$ dependence factorizes: it resides entirely in the kinematic coefficients which are clearly separated from the $\phi$-independent CFFs.
\end{enumerate} 
The $\phi$-dependent structure of $\sigma_{DVCS}$ and $\sigma_{\cal I}$, outlined above, is important in the method that we will employ to extract CFFs from DVCS data in the following Sections. An example of the various contributions to the cross section is shown in Figure \ref{fig:ExampleXUUPlot}.

\fixedfigure{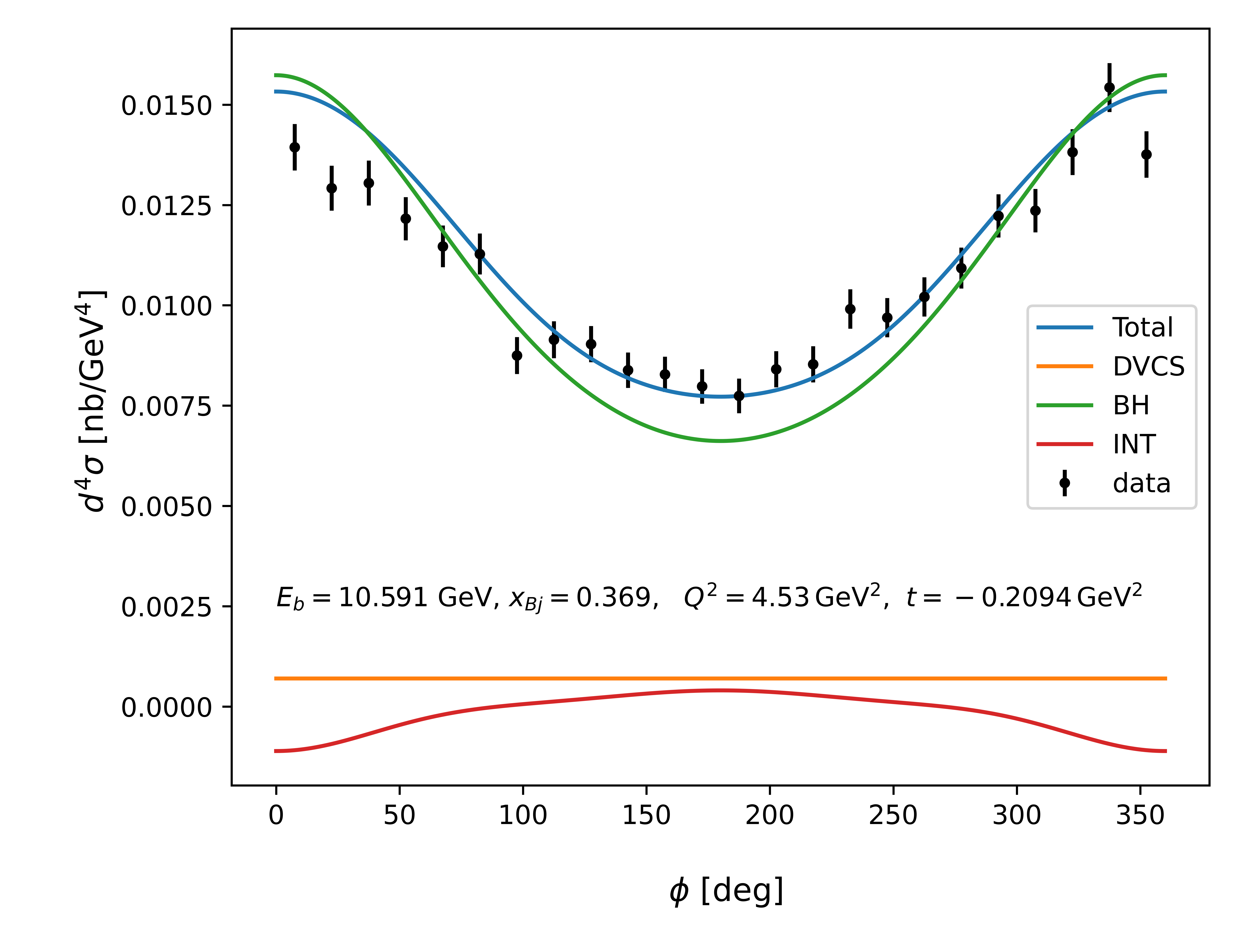}
{The expected relative contributions of the DVCS, BH, and interference terms to the total cross section, Eq.\eqref{eq:total-cross-section1}, in a particular kinematic bin. The DVCS and interference terms are calculated using the parametrization developed in \cite{Goldstein:2013gra, Kriesten:2021sqc}; experimental data from \cite{JeffersonLabHallA:2022pnx, Georges:2018kyi}.
In the plot the CFF values are fixed as:
ReH = -1.7927, 
ReE = -1.9572,
ReHt = 3.1316,
ReEt = 3.3615,
ImH = -1.9528,
ImE = -0.7543,
ImHt = -0.6698,
ImEt = -4.2128,
}
{fig:ExampleXUUPlot}

\vspace{0.3cm}
\noindent {\it Motivation}

\noindent In this paper we will not be concerned with the inverse problem of extracting GPDs from CFFs, but with the very first step of extracting  CFFs from experimental data.
As we explain below, the eight unknowns given by the real and imaginary parts of the proton CFFs (Eq.\eqref{eq:CFFdefn1}), can determine the shape of a curve fit through the cross section data at a given kinematic bin. 
These eight 
unknowns describing the cross section at leading order can vary considerably for each kinematic bin. 
In Figure \ref{fig:curvefits} we show, for illustration, what happens if one performs a curve fit allowing seven of them to vary but fixing $\Re e E$ to substantially different values chosen to be: $[-25, 0, 50]$. 
We can see how we obtain excellent fits for all three fixed values, with comparable $\chi^2$ per d.o.f. values. 
However, we can also show that once $\Re e E$ is fixed, $\Re e H$ and $\Re e \widetilde{H}$ become well constrained. 
It turns out that we are in the regime of bounded, but (statistically) covariant results for 3/8 twist-2 CFFs, which we present later
in Section \ref{sec:4}.
This result, which motivates our study, can be formalized using a likelihood analysis on each kinematic bin as explained in detail in Section \ref{sec:3}.

\fixedfigure{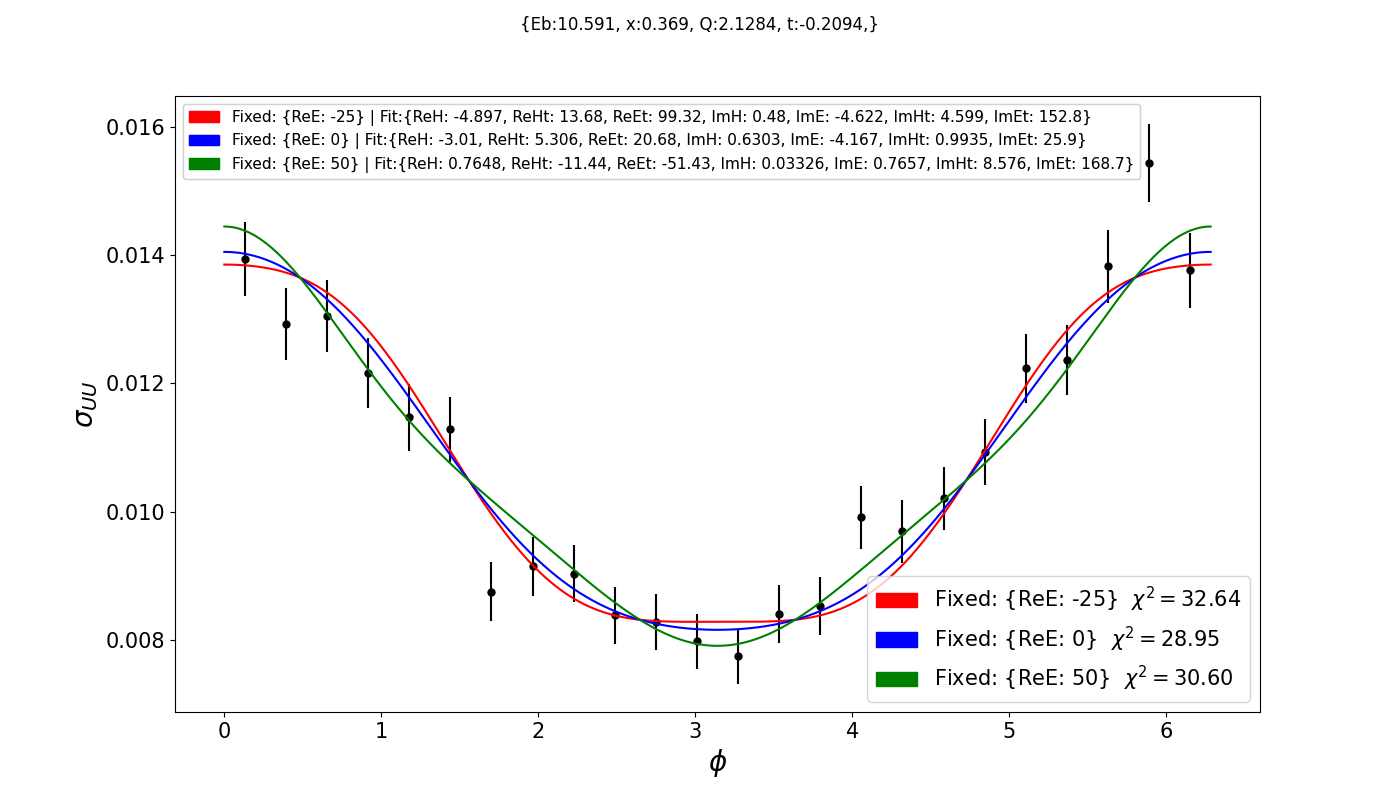}
{Illustration of how the DVCS cross section data ($\sigma_{UU}\equiv \sigma_{TOT}$ in Eq.\eqref{eq:total-cross-section1}, $\phi$ in radians)  allows good fits with wildly different CFFs.
For the particular kinematic bin shown here $\Re e E$ is set to each of [-25, 0, 50] and subsequently a fit is performed. 
The other seven twist-two CFFs are allowed to vary. 
The corresponding curve fits are all in excellent agreement with the data. 
Visually one can see the variation between the three curves is negligible compared to the width of the error bars on the data. 
Quantitatively we also confirm the negligible visual difference by calculating the $\chi^2$.
}
{fig:curvefits} 

Finally, we also provide a preliminary analysis including the extra sixteen Compton form factors which appear in cross section at twist-three \cite{Kriesten:2020wcx}. 
As we explain in what follows, the DVCS contribution in unpolarized scattering will no longer be independent of $\phi$. 
The impact of a twist-three analysis of  CFFs from DVCS data will be discussed in Section \ref{sec:4}.
Developing a strategy to include twist-three terms will be important as more data become available other approaches are given in Refs \cite{Belitsky:2001ns, Braun:2014sta, Belitsky:2010jw}, which can differ significantly beyond the leading order
(for a detailed discussion see \cite{Kriesten:2020wcx}).


\subsection{Experimental Data}
\label{sec:data}


For this analysis we have chosen to use the unpolarized DVCS data from the Jefferson Lab Hall A Collaboration \cite{JeffersonLabHallA:2022pnx, Georges:2018kyi}, 
with the kinematic parameters $[E, x_{Bj}, Q^2, t, \phi]$. For convenience, we are working with the dataset published on the Gepard GitHub page \cite{ GepardGithub} under the name BSS HALLA - 18. The reason for using this dataset is due to its binning; each bin has 24 datapoints of different $\phi$ with identical values of the other four parameters, which lends itself well to the analysis described in Section \ref{sec:3}. 

\begin{table}[h]
    \centering
    \begin{tabular}{|c|c|c|c|c|c|c|}
        \hline
        $E_{beam}$ (GeV) & $x_{Bj}$ & $Q^2$ $({\rm GeV}^2)$ & $t$ $({\rm GeV}^2)$ & $\phi$ (deg) & $\sigma_{total}$ & $\Delta \sigma$ \\
        \hline
        10.591 & 0.369 & 4.53 & -0.2094 & 7.5 & 0.01394 & 0.00058 \\
        \hline
        10.591 & 0.369 & 4.53 & -0.2094 & 22.5 & 0.01292 & 0.00056 \\
        \hline
        10.591 & 0.369 & 4.53 & -0.2094 & 37.5 & 0.01305 & 0.00056 \\
        \hline
        10.591 & 0.369 & 4.53 & -0.2094 & 52.5 & 0.01216 & 0.00054 \\
        \hline
        10.591 & 0.369 & 4.53 & -0.2094 & 67.5 & 0.01147 & 0.00052 \\
        \hline
        10.591 & 0.369 & 4.53 & -0.2094 & 82.5 & 0.01128 & 0.00051 \\
        \hline
        10.591 & 0.369 & 4.53 & -0.2094 & 97.5 & 0.00875 & 0.00046 \\
        \hline
        10.591 & 0.369 & 4.53 & -0.2094 & 112.5 & 0.00915 & 0.00046 \\
        \hline
        10.591 & 0.369 & 4.53 & -0.2094 & 127.5 & 0.00904 & 0.00045 \\
        \hline
        10.591 & 0.369 & 4.53 & -0.2094 & 142.5 & 0.00838 & 0.00044 \\
        \hline
        10.591 & 0.369 & 4.53 & -0.2094 & 157.5 & 0.00828 & 0.00044 \\
        \hline
        10.591 & 0.369 & 4.53 & -0.2094 & 172.5 & 0.00798 & 0.00043 \\
        \hline
        10.591 & 0.369 & 4.53 & -0.2094 & 187.5 & 0.00774 & 0.00043 \\
        \hline
        10.591 & 0.369 & 4.53 & -0.2094 & 202.5 & 0.00841 & 0.00045 \\
        \hline
        10.591 & 0.369 & 4.53 & -0.2094 & 217.5 & 0.00853 & 0.00045 \\
        \hline
        10.591 & 0.369 & 4.53 & -0.2094 & 232.5 & 0.00991 & 0.00049 \\
        \hline
        10.591 & 0.369 & 4.53 & -0.2094 & 247.5 & 0.00969 & 0.00049 \\
        \hline
        10.591 & 0.369 & 4.53 & -0.2094 & 262.5 & 0.01021 & 0.00049 \\
        \hline
        10.591 & 0.369 & 4.53 & -0.2094 & 277.5 & 0.01093 & 0.00051 \\
        \hline
        10.591 & 0.369 & 4.53 & -0.2094 & 292.5 & 0.01223 & 0.00054 \\
        \hline
        10.591 & 0.369 & 4.53 & -0.2094 & 307.5 & 0.01236 & 0.00054 \\
        \hline
        10.591 & 0.369 & 4.53 & -0.2094 & 322.5 & 0.01382 & 0.00057 \\
        \hline
        10.591 & 0.369 & 4.53 & -0.2094 & 337.5 & 0.01543 & 0.00061 \\
        \hline
        10.591 & 0.369 & 4.53 & -0.2094 & 352.5 & 0.01376 & 0.00058 \\
        \hline
    \end{tabular}
    \caption{Experimental data of the unpolarized DVCS cross section from Jefferson Lab Hall A \cite{JeffersonLabHallA:2022pnx, Georges:2018kyi} in 1 single kinematic bin used for our likelihood analysis.
    This table corresponds to various single bin likelihood analysis figures shown in this paper \ref{fig:corner_and_curve_canonical}
    \ref{fig:corner_and_curve_canonical_removedoutliers}
    \ref{fig:3CFF}.
    The full available dataset contains 44 such kinematic bins. 
    The summary of likelihood results for all kinematic bins is included in Table \protect\ref{tab:results} and Figure \ref{fig:Comp_Georges}.
    }
    \label{tab:data}
\end{table}

The data were taken after the Jefferson Lab beam energy upgrade to 12 GeV, and the energy spans the values $4.487\text{ GeV}\leq E\leq 10.992\text{ GeV}$. The other kinematic variables span the ranges $2.71\text{ GeV}^2\leq Q^2\leq 8.51\text{ GeV}^2$, $0.363\leq x_{\text{Bj}}\leq 0.617$ and $0.2094\text{ GeV}^2\leq -t\leq 1.3732\text{ GeV}^2$. For this kinematic regime the values of $\sqrt{-t/Q^2}$ range between 0.2 and 0.5, so we can expect some contamination from the twist-3 contribution. 

The published experimental data contains statistical uncertainties, as well as two types of systematic uncertainties, correlated and uncorrelated. The uncorrelated systematic uncertainties affect each experimental bin individually, while the correlated ones affect all bins equally. The uncorrelated uncertainties are ascribed to the contamination of background processes, most prominently semi-inclusive deep inelastic scattering (SIDIS), and the reported values of the uncertainties are between 2\% and 5\%. The correlated uncertainties come from several aspects of the analysis and the experimental setup, and their values range between 3\% and 6\%. These uncertainties depend on the kinematic bin, but also on the value of $\phi$. In all bins, the statistical uncertainties range between 3\% and 4\%, and their $\phi$ dependence is ignored. In this work we decided to only consider the statistical uncertainties because the systematic uncertainties have not been fully evaluated for this measurement. The authors report an average correlated uncertainty of 3\%, but, as mentioned above, estimate that it can go up to 6\%.
For more details, see \cite{Georges:2018kyi}.

\section{Deriving a likelihood function}
\label{sec:3}
We will use two different versions of a Bayesian likelihood analysis to place bounds upon the Compton form factors (CFFs). 
We denote the first version as ``canonical" as it is simpler as we can ultimately write the total joint likelihood of the parameters of interest as a simple product of Gaussians.
The second version is more complicated as it involves taking the differences of two total cross sections at different $\phi$ angles, and calculating a corresponding covariance matrix for the differences calculated.
The first version is easier to understand but the second method provides tighter bounds on the CFFs. 
The reason for the tighter bounds can be intuitively understood by realizing the same information is used in both versions, but the first allows four parameters to vary, while the second only has three parameters. 
A model with fewer parameters can  be more tightly constrained using the same information as another with more parameters. 
It is then illustrative to show that both methods provide similar results and we can view the simpler method as a reasonable sanity check to the more complicated method. 

For both versions, {we assume that the data can be randomly selected from a normal distribution}. If the model of the cross section was perfect and the measurements had no error we would expect 
\[\sigma_{model}(\phi) = \sigma_{exp}(\phi) \]
However,  the observations do have error thus we can view each cross section measurement as independent and sampled from a true value calculated from the model $\mu = \sigma_{model}$. 

In what follows  we give a detailed description of each of the two likelihood analysis approaches, and show their results in Section \ref{sec:4}.

\subsection{Canonical Likelihood}
\label{sec:simple}
The canonical method's likelihood can be written as a product of gaussians.  
For a fixed set of kinematic parameters $[E, x_{Bj}, Q^2, t]$, the total cross section is measured at 24 azimuthal $\phi$ angles (we show an example of the dataset in Table \ref{tab:data}), and can be also separately calculated using a phenomenological model. In this work we use the model from Ref.\cite{Kriesten:2019jep,Kriesten:2020wcx}.
Assuming our model is correct, the calculated cross sections will still differ from the observed cross sections. 
The difference between model and observation must be explained by the error bars (measurement error).
Such error bars are reported by the experiment and are available in the observed data table from \cite{GepardGithub}, discussed in section \ref{sec:data}.
For a single datapoint (one $\phi$ value), the probability that the model differs from an observation is described by a univariate Gaussian probability density function.
Thus using such a univariate Gaussian, we compare cross section observed ($ d^4 \sigma_{exp} \equiv x$) against the model value expected ($ d^4 \sigma_{model} \equiv \mu$), and use the error bar (standard deviation $\sigma$) to calculate the probability of a single row of our data table assuming the model is correct.

We can look at 24 rows of the data table simultaneously which should have the same CFFs values, same kinematic coefficients $[E, x_{Bj}, Q^2, t]$, but different values of $\phi$.
The probability of seeing such 24 rows (of different $\phi$ values) is described as the product of 24 single row univariate gaussians.
The total likelihood of the CFFs for fixed $[E, x_{Bj}, Q^2, t]$ can then be written,
\begin{equation}
    \begin{aligned}
        \mathcal{L}_{canonical}( {\rm CFFs}) = 
        \prod_{i=1} ^ {24}
       {\rm  Gaussian} ( x = d^4\sigma_{exp}(\phi_{i}), \mu= d^4\sigma_{model}(\phi_{i}), \sigma = Err(d^4 \sigma_{exp, i}) ) ,
    \end{aligned}
\end{equation}
where,
\begin{equation}
    \begin{aligned}
        {\rm Gaussian}(x, \mu, \sigma) =  \frac{1}{\sqrt{2\pi \sigma^2}}\exp\bigg[ -\frac{(x - \mu)^2}{ 2\sigma^2} \bigg] 
         \propto  \text {exp}\bigg[ - \frac{1}{2} \left( \frac{x - \mu}{ \sigma} \right)^2 \bigg],
    \end{aligned}
\end{equation}

Note that we can strategically omit the normalization in the front of the gaussian $\frac{1}{\sqrt{2\pi \sigma^2}}$ because there is no dependence on the parameters. 
Thus by omitting it, only the normalization is lost, and the computational speed will improve for the Markov chain Monte Carlo (MCMC) method that we implement below.

\subsection{Difference Likelihood}
\label{sec:difference}
The more constraining method involves calculating some differences of cross sections for combinations choices of two angles. We will denote two rows of the data table as row $A $ and row $B$ accordingly. Each one of the rows, $A$ and $B$, have  angles ($\phi_{A}, \phi_{B}$) and cross sections ($d^4\sigma_{exp, A} \equiv x_A$ , $d^4 \sigma_{exp, B} \equiv x_B $).
We wish to calculate the probability of seeing the difference of two cross sections. 

It is helpful to define a difference of two cross sections for both the model, ($d^4\sigma_{model, A} \equiv \mu_A$, $d^4\sigma_{model, B} \equiv \mu_B$) besides the one for the data,
\begin{equation}
    \begin{aligned}
        \mu_{A,B} & =  \mu_A - \mu_B \\
        x_{A,B} & =  x_A - x_B
    \end{aligned}
\end{equation}

Our data table contains 24 possible choices for each of $A$, $B$. 
Each choice also has the potential to be correlated by each other choice because the same information may have been used. 
Let one choice of $[A, B]$ be defined as $[A_{i}, B_{i}]$, and another choice as $[A_{j}, B_{j}]$, where $[A_{i}, B_{i}, A_{j}, B_{j}]$ can each be a different choice of row index between $[0, 24]$, for a total of 4 possibly different indexes and $24^4$ possible combinations. For example, $A_{i}=\sigma_{row0}, 
B_{i}=\sigma_{row1}, 
A_{j}=\sigma_{row2}, 
B_{j}=\sigma_{row3}$ is one such possibility corresponding to the first 4 rows of the data table. 
There are $24^4$ such possibilities.
The covariance between two such difference-pairs of two can be calculated using the algebra of Gaussian random variables as follows,

\begin{equation}
    \begin{aligned}
        \label{eqn:crosssection_difference_covariance}
        cov( x_{A_i,B_i} ,  x_{A_j,B_j} ) 
        &=  cov( 
        x_{A_i} - 
        x_{B_i},  
        x_{A_j} - 
        x_{B_j} ) \\ 
        &= cov(  x_{A_i}, x_{A_j} ) 
        - cov(  x_{A_i}, x_{B_j} ) 
         - cov( x_{B_i} , x_{A_j} )
        + cov( x_{B_i}, x_{B_j} ) ,
    \end{aligned}
\end{equation}
where the covariance of two measurements $cov(x_{\alpha}, x_{\beta})$ is either zero, or the regular variance of the same measurement, {\it i.e.}  we can represent the covariance with the following Kronecker delta relationship,
\begin{equation}
    \begin{aligned}
        cov( x_{\alpha}, x_{\beta} ) = 
        \delta_{\alpha, \beta}  \, (\sigma_{x_{\alpha}})^2
    \end{aligned}
\end{equation}
The covariance between any two choices of cross section differences, Eq.\eqref{eqn:crosssection_difference_covariance}, can be expressed as a covariance matrix. If we used all possible angle differences the covariance would be of shape $(576 \times 576)$. 
However in practice we can only use 23 difference angles, where we subtract the cross section at the same angle from the cross section at the other 23 angles for a covariance matrix of shape $(23 \times 23)$. 
We can see by checking the rank of the covariance matrix that the rows are not linearly independent if we use all possible angle differences. All the available information is used if we use only 23 angle differences. 

The observed value for the difference of cross sections is a simple subtraction. 
There will be 23 such subtractions performed creating a vector with 23 elements:

\begin{equation}
    \begin{aligned}
        \mathbf{x}= \begin{bmatrix}
        x_{\phi_0} - x_{\phi_{23}} \\
        x_{\phi_1} - x_{\phi_{23}} \\
        ... \\
        x_{\phi_{22}} - x_{\phi_{23}}
        \end{bmatrix}
    \end{aligned}
\end{equation}

Given that we are assuming the cross section model is true, the mean values we expect to observe are what the cross section model predicts:

\begin{equation}
    \begin{aligned}
        {\bf \mu} = \begin{bmatrix}
        \mu_{\phi_0} - \mu_{\phi_{23}} \\
        \mu_{\phi_1} - \mu_{\phi_{23}} \\
        ...  \\
        \mu_{\phi_{22}} - \mu_{\phi_{23}}
        \end{bmatrix}
    \end{aligned}
\end{equation}
The 23-dimensional covariance matrix can now be used in a dimensional multivariate Gaussian to define the total un-normalized likelihood,
\begin{equation}
    \label{eqn:likelihood_crosssectionDIFF}
    \begin{aligned}
        \mathcal{L} = \text{Gaussian}( x \equiv \sigma_{exp,A} - \sigma_{exp,B} , \mu \equiv  \sigma_{model,A} - \sigma_{model,B}, \Sigma = cov(\sigma_{exp,A_i,B_i} , \sigma_{exp,A_j,B_j} )  ), \quad \quad \forall A, B 
    \end{aligned}
\end{equation}
Where here ``Gaussian" represents the multivariate Gaussian probability density function,
\begin{equation}
    \label{eqn:likelihood_crosssectionDIFF}
    \begin{aligned}
        \text{Gaussian}( x, \mu, \Sigma)  = \frac{1}{2 \pi |\Sigma|} \text{exp}\bigg( -\frac{1}{2} (x-\mu)^T \Sigma^{-1} (x-\mu)  \bigg)
    \end{aligned}
\end{equation} 

Note that the multivariate Gaussian reduces to (1) the univariate Gaussian when the covariance matrix is $1\times1$ or (2) a product of univariate Gaussians when the covariance has all 0 off-diagonal elements. 
In our case we use 23 possible differences of observed cross sections yielding a covariance matrix which is $23\times23$ with non-zero off-diagonal terms which does not reduce further. 

\subsection{Bayesian Markov Chain Monte Carlo (MCMC)}
For both the canonical and the difference likelihood methods derived above, we need to make few assumptions in the Bayesian data analysis paradigm to proceed. 
{Bayes’ theorem notably allows us to write the conditional probability of A, given that B is true, $P(model|data)$, or the posterior, as the product of B given that A is true, $P(data|model) = $ the likelihood, multiplied by the ratio of the single probabilities of A, $P(model)$, the prior, and B, $P(B)$, the evidence. 
In our case, A is our cross section model, and B is the set of experimental data.
In order to calculate the posterior probability (model inference), or the best parameters in our cross section model (the CFFs) to reproduce the data, we need to set the prior and evidence probability distributions.}
We will use the uniform prior on the interval $(-\infty, \infty)$ for the CFF values because we have no non-data-based constraints on the probability of what the CFFs should be. 
Setting the value of the prior probability to $1$ (one) is a numerically practical way to admit complete ignorance of parameters in our {uninformative prior}.
We also assume the probability of evidence obtained to also be one. 
Then, our likelihood function can be treated as an un-normalized improper posterior, and the parameters can be subsequently explored.
For the analysis of DVCS data at hand, this will result in considering,
\begin{equation}
    \label{eqn:posterior}
    \begin{aligned}
        Posterior \bigg( 
        CFFs
        \bigg) 
        \approx \frac{Likelihood \times Prior}{Evidence}
        \approx \frac{ Likelihood \times 1 }{ 1 } 
    \end{aligned}
\end{equation}
Next, Markov Chain Monte Carlo (MCMC) algorithms are used to take multidimensional probability density functions and generate a set of representative samples that obey and represent the underlying distribution. 

\subsubsection{Markov Chain Monte Carlo (MCMC) Sampling}
Markov Chain Monte Carlo (MCMC) sampling is the procedure of starting with a probability density function $\mathcal{L}$ of multiple parameters $ \overrightarrow{\theta}$ and attempting to generate a set of samples of the parameters which represents the distribution ({\it i.e.}  table of $\overrightarrow{\theta}$ values). 
There are a few primary reasons one generates such samples:

\noindent (1) To marginalize over less interesting parameters without the requirement to integrate a closed form symbolic expression for the probability; 

\noindent (2) to use the samples in further calculations {\it i.e. } substitute each row of generated samples into some other function of the samples which returns a new number;

\noindent (3) to create easy to use visualizations of the samples which we can look at to discern our own human belief in the parameters. 

Here, we have used the results of MCMC in items (1) and (3). 
We have not done (2) and implemented the MCMC samples into further calculations which depend upon them. 
However, as the hadronic physics community has great interest in generalized parton distribution functions (GPDs), a natural extension of our analysis will be to use the table of CFFs samples and compare them to various models of GPDs.
One should be able to show preference for some GPD models over others and potentially be able to rule out some classes of such models. 

Applying MCMC to likelihood functions is common practice. One can use the generated samples to visualize the underlying distribution. 
One can also use the samples as a way of numerically marginalizing over parameters which are of lesser interest (nuisance parameters). 
{For instance, in our case, $\sigma_{DVCS}$ in the canonical approach is a nuisance parameter as we only focus on the three extractable Compton form factors.}
Furthermore,  we use the samples for visualization purposes. We also use the samples to approximate results as a multivariate Gaussian. 

There is a vast literature covering the various approaches used to generate samples. We adopted the 
\texttt{emcee} ensemble sampling algorithm.
In a nutshell, an MCMC will start from a given position in parameter space where the posterior probability is known, take a step in parameter space randomly, and then evaluate the posterior probability at the new point; subsequently, a random number is compared with the ratio of the calculated posterior to the previous probability: if the ratio is less than one, the step is accepted and the new parameters are added to the chain, while if it is greater than one, it is rejected and the previous parameters are kept in the chain. 
The most basic such algorithm is the Metropolis-Hastings algorithm.
Many MCMC algorithms exist which are more sophisticated and obtain samples more efficiently with faster convergence.
The focus of the work presented here is on showing what an MCMC does, and how the resulting samples can be interpreted/used. 
A detailed explanation is well beyond the scope of this paper, and we direct the reader to the method we have chosen to use \cite{Foreman_Mackey_2013,Foreman-Mackey2016}.
{(a combination of the citations in these papers as well as Wikipedia will provide an excellent introduction and deep explanation to the inner workings of MCMC algorithms). }

In the next Section we show results for the extraction of CFFs from data based on this framework,
{{\it i.e.} based on the likelihood function which describes the probability of CFFs, running the MCMC with the corresponding posterior probability distributions to generate samples.}

\section{Results}
\label{sec:4}
In what follows, both versions of the likelihood analysis, ``canonical" (Sec.\ref{sec:simple}) and ``difference" (Sec.\ref{sec:difference}), defined above are applied to DVCS data. Various specific issues arise in the numerical analysis that are discussed in anticipation of more and varied data in this sector, including polarized scattering and timelike Compton scattering (TCS) in kinematic settings from Jefferson Lab to the EIC.  
In exclusive photon electroproduction, it is apparent, from the formulation of the leading order cross section in terms of CFFs (Section \ref{sec:2}), that only three CFFs, specifically the ones defining the interference term, Eq.\eqref{eq:sigmaINT1}, can be non-degenerate. 

\subsection{Canonical}
\label{sec:simple_results}
In the canonical likelihood analysis an additional parameter is introduced into the cross section that accounts for the degeneracy of five CFFs. The latter is chosen as the DVCS cross section which, as one can see from Eq.\eqref{eq:sigmaDVCS1}, is both independent from $\phi$, and it contains all eight CFFs, therefore, even if it includes the variable parameters, $\Re e {\cal H}$, $\Re e {\cal E}$, $\Re  e \widetilde{{\cal H}}$, it can, nevertheless, be used as an additional, independent parameter. 

We present results of the posterior distribution from the MCMC for the $3+1$ parameters in this model, in the form of a corner plot which best displays the correlations (plots were obtained using the corner.py library) \cite{Foreman-Mackey2016}. 
From this analysis we draw the conclusion that all three non degenerate CFFs are highly correlated with one another. Their values can only be extracted with large error bars. 
The mean values of the form factors, along with the $\sigma$ values are shown in Table \ref{tab:canonical_singlebin_outlier_compare}.

\begin{figure}[h!]
    \includegraphics[width=8cm]
    {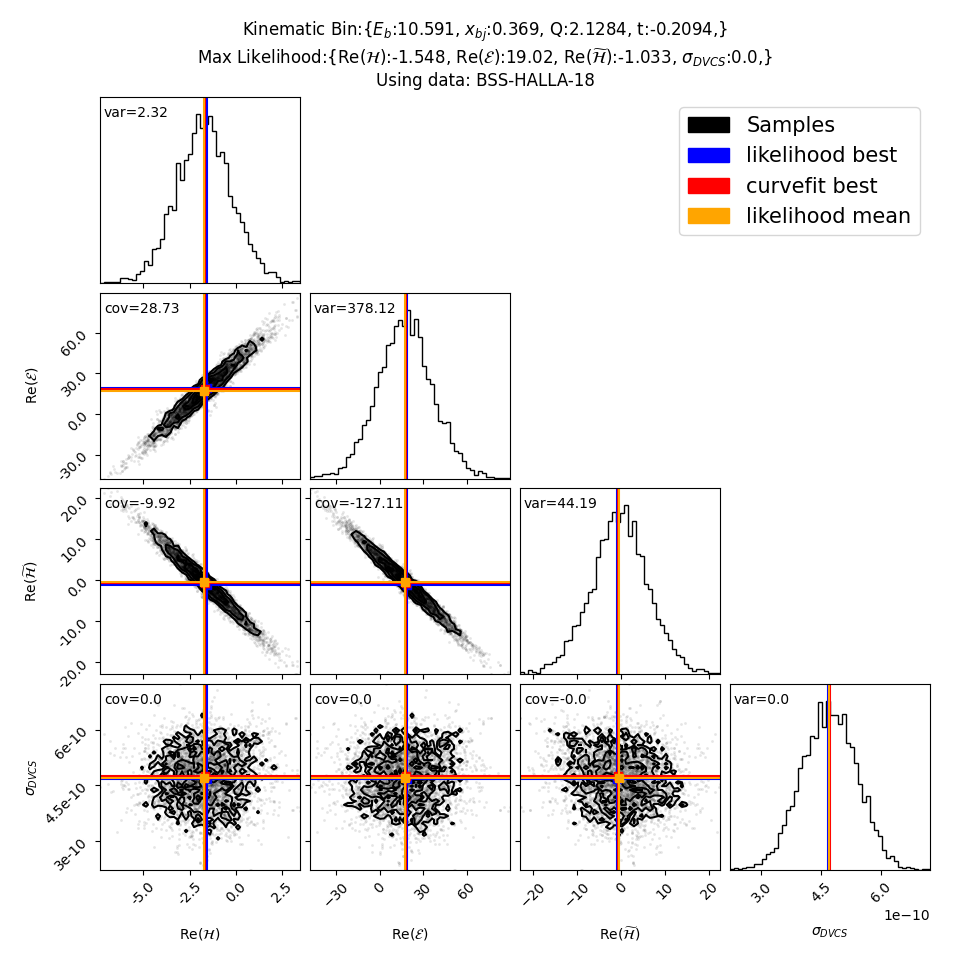}
    \hspace{1cm}
    \includegraphics[width=8cm]
    {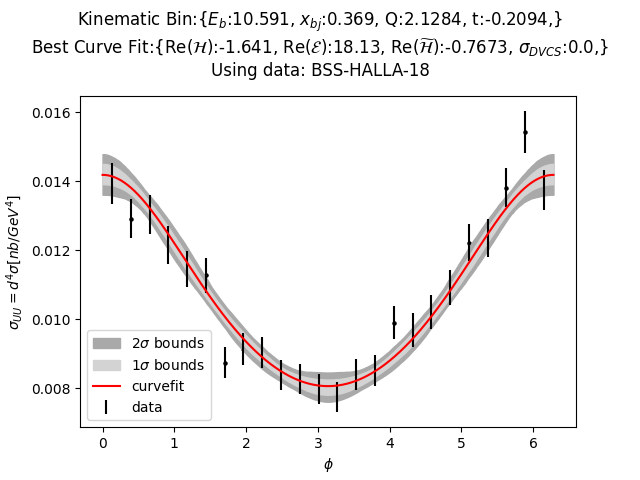}
\caption{Canonical Result. Left: Corner plot using corner.py \protect\cite{Foreman-Mackey2016} of three CFFs: $\Re e {\cal H}$, $\Re e {\cal E}$, $\Re  e \widetilde{{\cal H}}$, as well as an extra parameter, $\sigma_{DVCS}$, treated as a fourth free parameter independent of the angle $\phi$.
    Right: Error bands are generated by drawing 1000 curves, and then at each $\phi$ value finding the cross section values which is located above/below 95\% of the curves.
    The Red curve corresponds to the best fit CFF parameters using a global optimizer geared towards general least squares in python scipy.
    The Blue curve corresponds to the MCMC sample of CFF parameters which has the largest likelihood.
    Red and Blue curves have nearly identical CFF parameters. 
}
\label{fig:corner_and_curve_canonical}
\end{figure}

\clearpage
\textcolor{blue}{To investigate the situation of highly correlated results} we performed \textcolor{blue}{an illustration} of the dependence on the experimental data distributions, focusing, in particular, on  the following questions:
\begin{enumerate}
    \item if outliers can be identified, and subtracted, does this lead to an improved analysis?
    \item are the data too noisy?
\end{enumerate}

\begin{table}
    \centering
    \begin{tabular}{|c|c|c|c|c|c|c|c|c|c|}
    \hline
         Model &   $\mu_{H}$ & $\mu_{E}$ & $\mu_{Ht}$ &
         $\Sigma_{H,H}$ & 
         $\Sigma_{E,E}$ & 
         $\Sigma_{Ht,Ht}$ & 
         $\Sigma_{H,E}$ &  
         $\Sigma_{E,Ht}$ & 
         $\Sigma_{H,Ht}$  \\
         \hline 
     Canonical &  -1.69  & 17.6 &  -0.549   
            &  2.19 & 371 & 41.6 & 27.7 & 122 & 9.36 \\
         \hline
      No outls.   
            &  -1.251   & 25.18 &     -4.146
            &   2.28   & 367 &  43.2 & 27.9 & 123 &  9.67 \\        
    \hline
      \end{tabular}
\caption{Table of single kinematic bin results.  CFF results for the canonical fit. 
    The CFFs shown are the mean value as well as their covariances.
    Results for different methods of outlier testing are shown.
    }
\label{tab:canonical_singlebin_outlier_compare}
\end{table}

\begin{figure}[h!]
    \includegraphics[width=8cm]
    {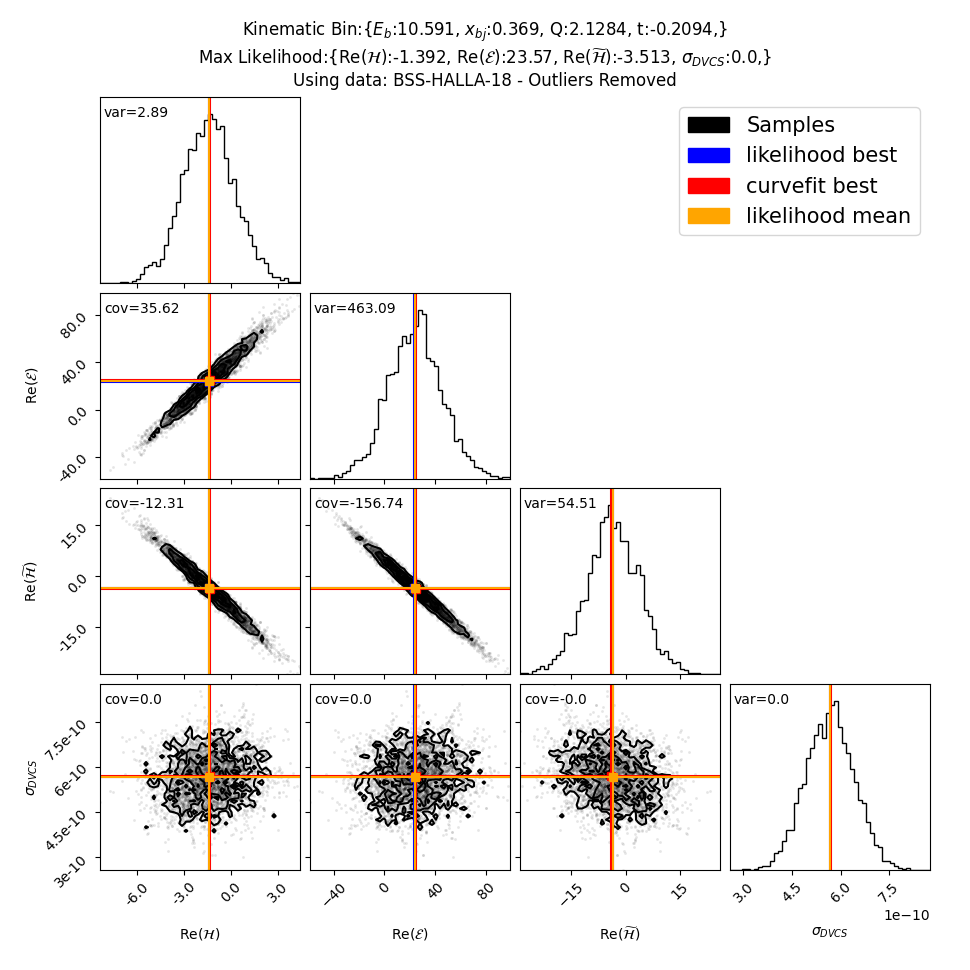}
    \hspace{.1cm}
    \includegraphics[width=8cm]
    {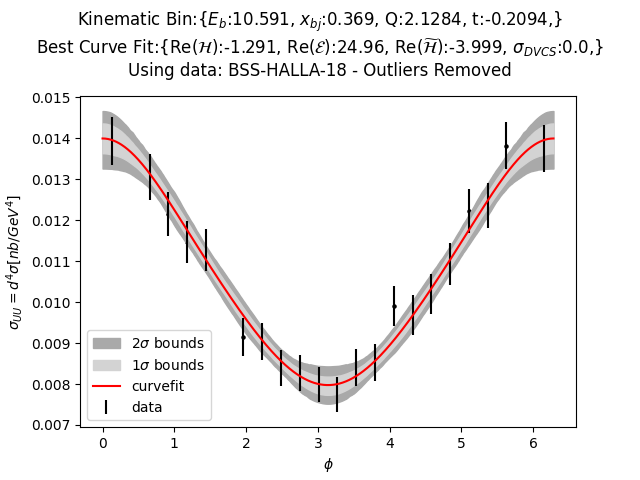}
    \caption{Left: Corner plot resulting from the analysis where experimental data which are considered to be outliers are removed from the analysis. The latter are shown plotted vs. $\phi$; Right: Graph of DVCS cross section with outlier points removed. 
    On the right plot error bands are generated by drawing 1000 curves, and then at each $\phi$ value finding the cross section values which is located above/below 68\% or 95\% of the curves.
    The Red curve corresponds to the best fit CFF parameters using a global optimizer geared towards general least squares in python scipy.
    The Blue curve corresponds to the MCMC sample of CFF parameters which has the largest likelihood.
    Red and Blue curves have nearly identical CFF parameters. 
    }
    \label{fig:corner_and_curve_canonical_removedoutliers}
\end{figure}

The outcome of point 1) is reported  in Figure \ref{fig:corner_and_curve_canonical_removedoutliers} where on the {\it lhs} we show a corner plot that correspond to the data once the outliers are removed as in the graph shown on the figure's {\it rhs}. 
Results of the central values of the CFFs and details of the error analysis are  presented in Table \ref{tab:canonical_singlebin_outlier_compare} (second  row).

We notice that, while the accuracy of the parameters prediction seems to improve, leading to a softening of the strong correlation found in the initial analysis, the treatment of outliers considered here is not founded on an objective methodology. Outliers were chosen expected from the symmetry of the $\phi$ dependence and removed points were those that lay at least $2\sigma$ away from the curve.  Nevertheless, the present experimental data does show points that are substantially farther from the expected symmetric behavior in $\phi$, granting the necessity of future development of a dedicated procedure on how to systematically and properly treat this issue quantitatively as more data sets become available. 

We also performed the fit including only half of the data points for each $\phi$ spectrum, that is choosing either the points with $\phi \equiv [0,\pi]$ or $\phi \equiv [\pi,2\pi]$. The reason to do this is that, while reducing the number of available data points, this method provides yet another way to overcome the potentially problematic question of the lack of symmetry in $\phi$ of experimental data. Nevertheless, the results show very little variation from the original plots in Fig.(\ref{fig:corner_and_curve_canonical}), indicating that the deviations of the data from a perfectly symmetric distribution around $\phi=\pi$, do not affect the outcome. 


\clearpage
\subsection{Difference}
\label{sec:Difference_results}
\begin{figure}[h!]
    \includegraphics[width=8cm]
    {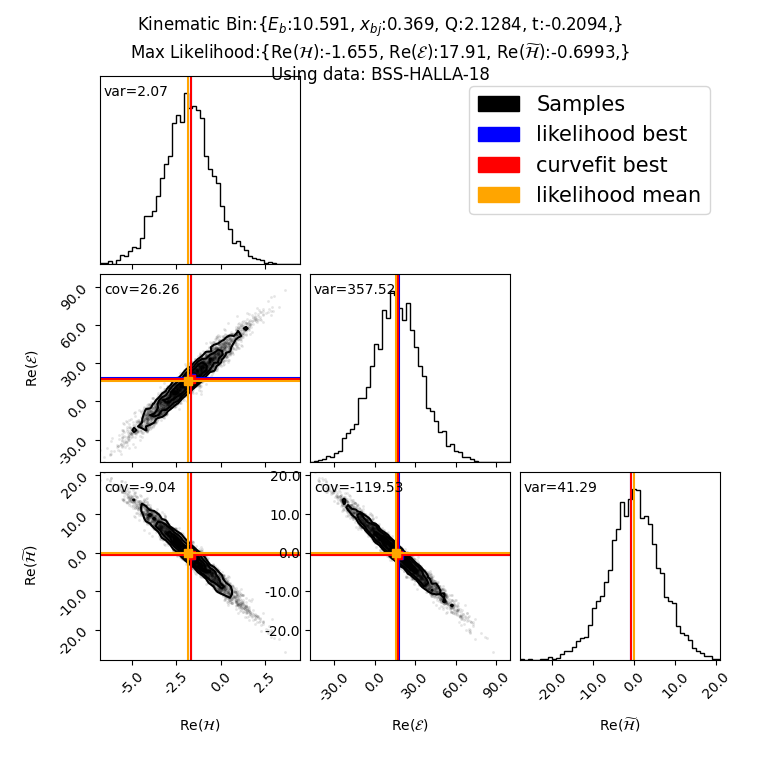}
    \hspace{0.1cm}
    \includegraphics[width=8cm]
    {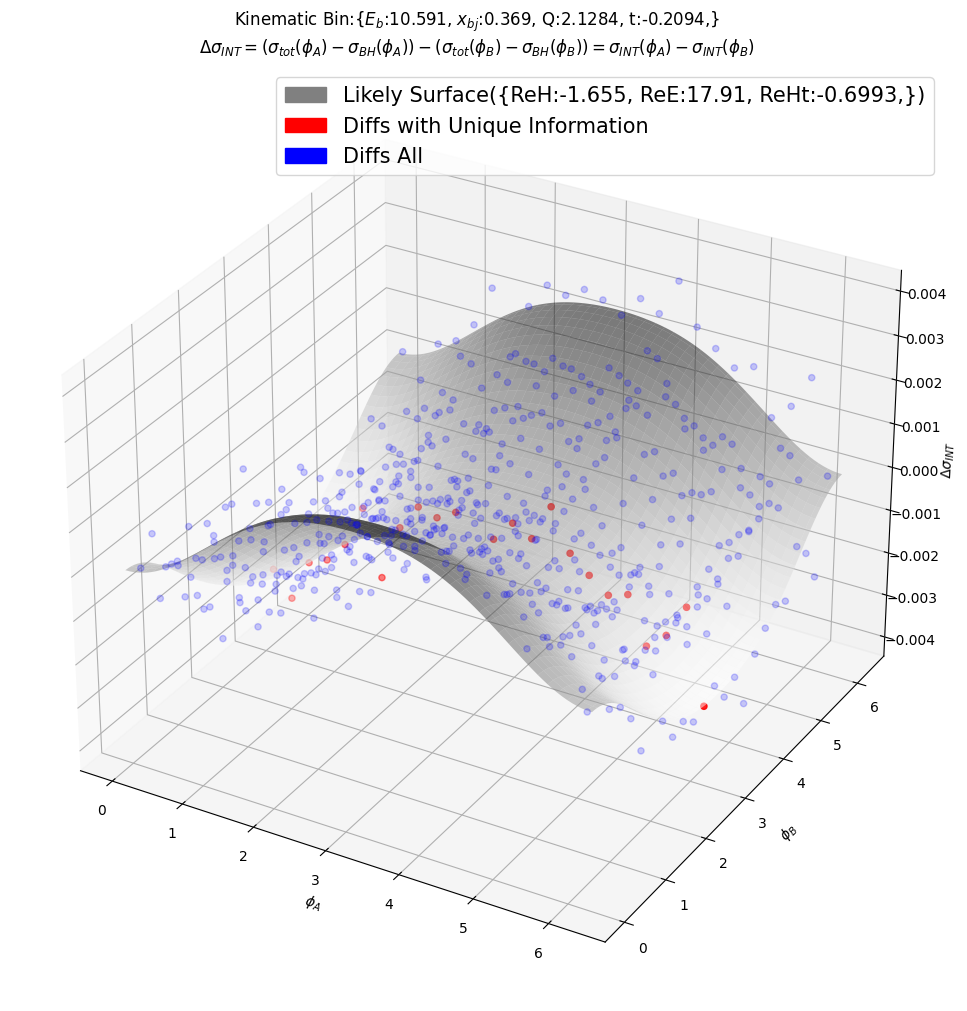}
    \caption{ 
    Left:
    Corner plot using the corner.py library  of 3 CFFs: $\Re e {\cal H}$, $\Re e {\cal E}$, $\Re e \widetilde{\cal{H}}$. 
    The cross section difference likelihood is defined as in equation \ref{eqn:likelihood_crosssectionDIFF}.
    The cross section difference likelihood is explored using the emcee MCMC algorithm \cite{Foreman_Mackey_2013}. 
    Right: Surface plot of the maximum likelihood parameters against the difference of two cross sections at all possible angles. 
    The Blue points are all possible cross section differences. 
    The Red points are those used in performing the fit to obtain the grey surface. 
    The information contained in all blue points is fully contained in the red points only. 
    }
    \label{fig:3CFF}
\end{figure}

Next, we show results of a likelihood analysis which assesses the difference of two cross section measurements at different angles (Figure \ref{fig:3CFF}). 
The difference method uses a multivariate gaussian with a covariance matrix. 

\begin{figure}[h!]
    \includegraphics[width=8cm]{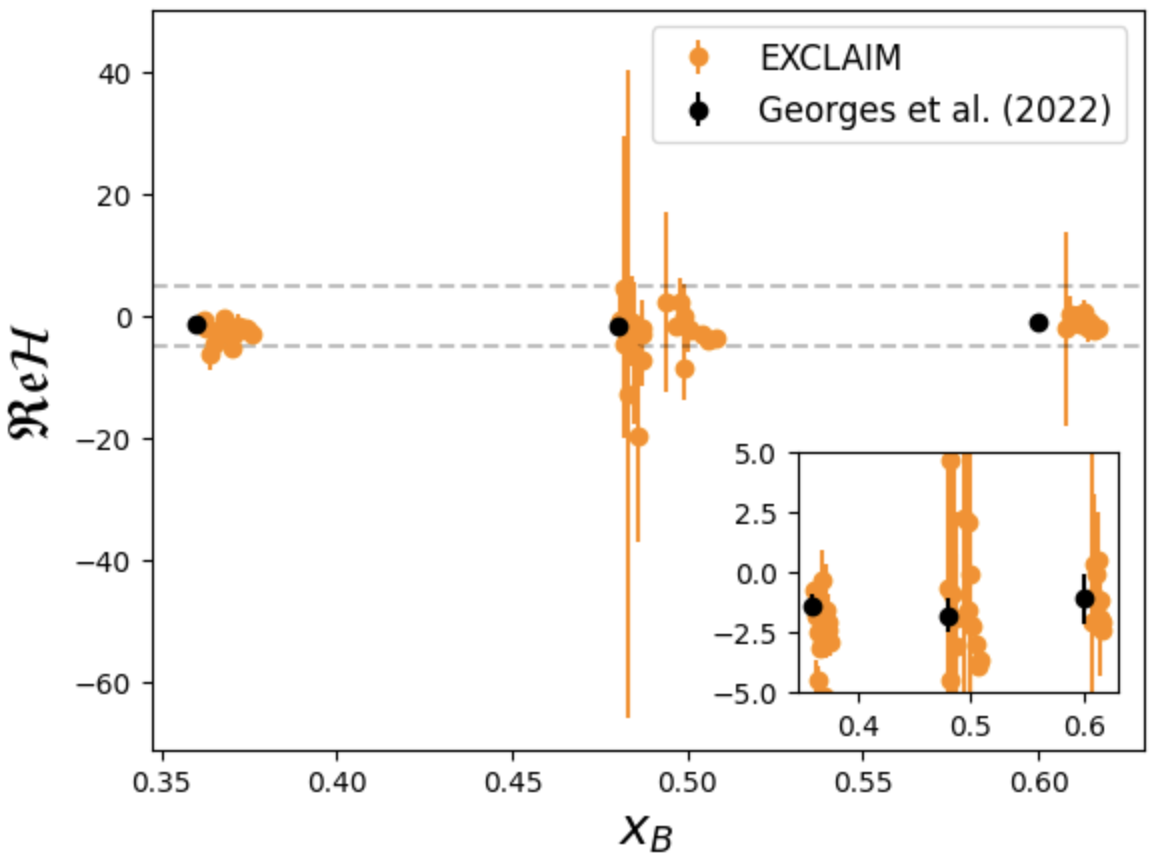}
    \hspace{0.1cm}
    \includegraphics[width=8cm]{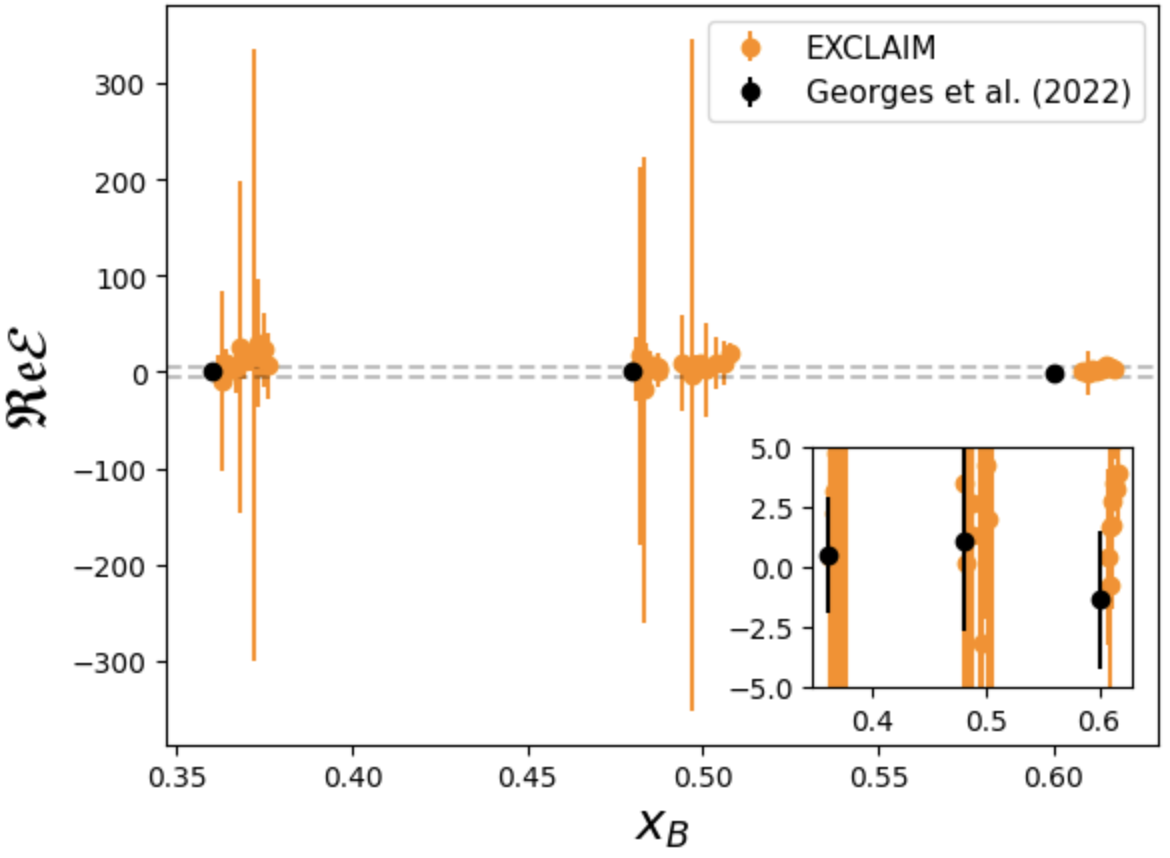}
    \includegraphics[width=8cm]{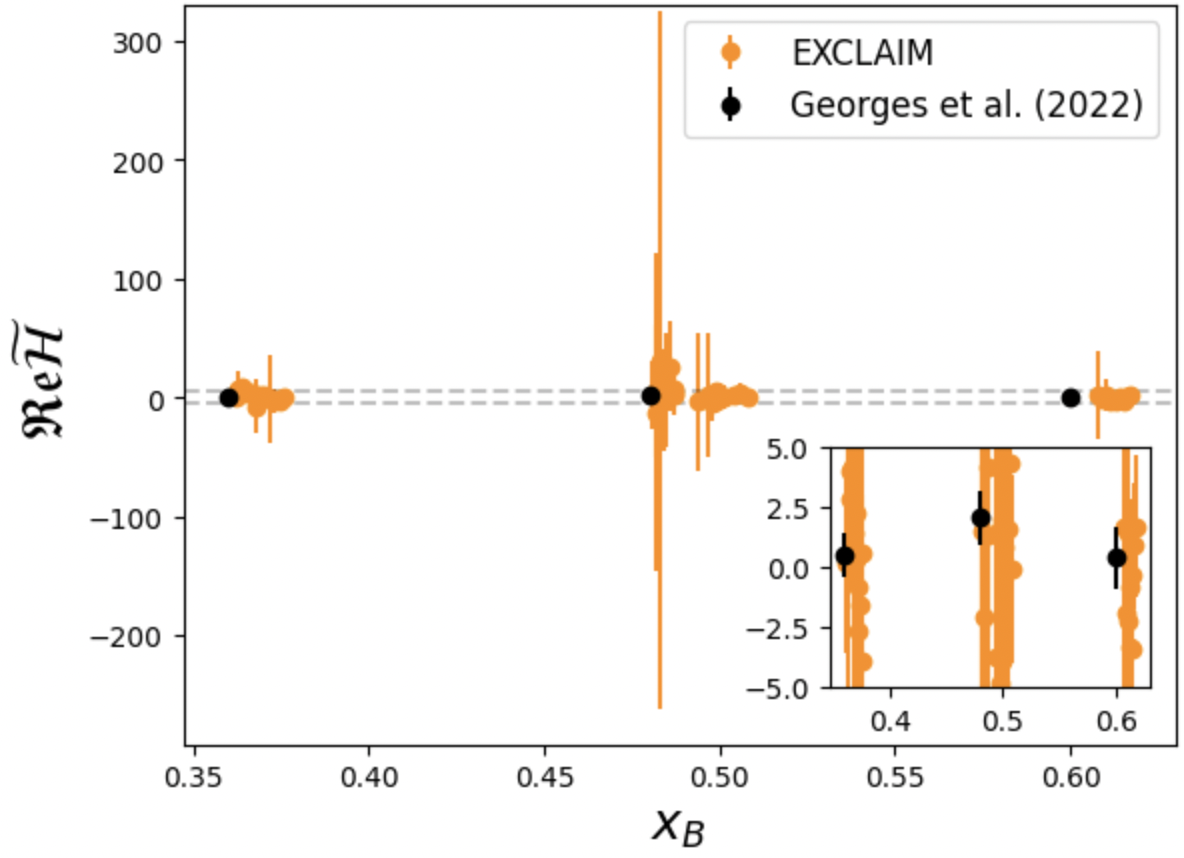}
    \caption{ 
    Visualization of the Compton form factors presented in Table \ref{tab:results} for the three kinematic bins in $x_{B}$ reported in \cite{JeffersonLabHallA:2022pnx} compared to the CFFs extracted in \cite{JeffersonLabHallA:2022pnx}. The EXCLAIM error values correspond to the covariances $\Sigma_{H, H}$, $\Sigma_{E, E}$, and $\Sigma_{\widetilde{H},\widetilde{H}}$. 
    }
    \label{fig:Comp_Georges}
\end{figure}

In this case, the DVCS cross section drops out of the analysis. The numerical values of the extracted parameters are reported in Table \ref{tab:results}. 
\textcolor{blue}{Further, we show the comparison of these obtained Compton form factors in Figure \ref{fig:Comp_Georges} for the three kinematic bins in $x_{B}=0.36, 0.48, 0.6$ reported in \cite{JeffersonLabHallA:2022pnx} with the CFFs extracted in \cite{JeffersonLabHallA:2022pnx}. The EXCLAIM error values correspond to the covariances $\Sigma_{H, H}$, $\Sigma_{E, E}$, and $\Sigma_{\widetilde{H},\widetilde{H}}$.
}

\begin{table}[h!]
    \centering
\begin{tabular}{r|r|r|r|r|r|r|r|r|r|r|r|r}
\hline
    $E$ (GeV) & \ \ \ \  $x_{Bj}$ & $Q$ (GeV)  & $t$ (GeV$^2$) & $\mu_{H}$ & $\mu_{E}$ & $\mu_{Ht}$&  
    $\Sigma_{H,H}$ &  
    $\Sigma_{E,E}$  &  
    $\Sigma_{Ht,Ht}$ &  
    $\Sigma_{H,E}$ &  
    $\Sigma_{H,Ht}$ &  
    $\Sigma_{E,Ht}$ \\
\hline
 4.487 & 0.483 & 1.646 & -0.391 &  -4.493 &   0.173 &   5.454 &         9.726 &        35.722 &          55.462 &        18.007 &        -22.596 &        -42.475 \\
 4.487 & 0.483 & 1.646 & -0.348 & -12.779 & -18.875 &  30.908 &        52.929 &       242.426 &         293.722 &       112.673 &       -123.436 &       -263.223 \\
 4.487 & 0.484 & 1.646 & -0.435 &  -0.927 &   8.229 &  -2.092 &         7.424 &        20.815 &          42.313 &        11.453 &        -17.004 &        -27.473 \\
 4.487 & 0.485 & 1.646 & -0.480 &  -6.142 &   5.385 &   6.293 &        11.723 &        17.263 &          47.518 &        11.581 &        -22.028 &        -25.109 \\
 4.487 & 0.485 & 1.649 & -0.540 & -19.579 &   6.592 &  26.070 &        17.562 &         8.300 &          37.734 &         3.584 &        -22.864 &        -10.353 \\
 7.383 & 0.363 & 1.780 & -0.297 &  -0.776 &   0.393 &   0.186 &         0.197 &        16.700 &           3.797 &         1.284 &         -0.743 &         -7.005 \\
 7.383 & 0.363 & 1.780 & -0.211 &  -1.867 &  -9.750 &   6.738 &         0.890 &        94.007 &          14.793 &         8.699 &         -3.503 &        -36.194 \\
 7.383 & 0.364 & 1.783 & -0.586 &  -6.223 &   9.507 &   9.146 &         2.535 &        14.261 &           6.156 &        -2.366 &         -2.234 &         -3.500 \\
 7.383 & 0.365 & 1.783 & -0.471 &  -4.490 &   2.274 &   4.029 &         0.569 &        12.488 &           4.515 &         0.540 &         -0.969 &         -5.578 \\
 7.383 & 0.365 & 1.783 & -0.385 &  -2.467 &   3.188 &   2.867 &         0.200 &        11.356 &           3.134 &         0.533 &         -0.544 &         -4.787 \\
 8.521 & 0.367 & 1.910 & -0.266 &  -3.154 &   4.729 &   4.138 &         0.256 &        27.333 &           5.037 &         2.154 &         -1.029 &        -10.802 \\
 8.521 & 0.367 & 1.910 & -0.205 &  -0.324 &  24.985 &  -7.389 &         1.256 &       172.650 &          21.984 &        14.256 &         -5.124 &        -60.197 \\
 8.521 & 0.369 & 1.916 & -0.330 &  -2.951 &  10.543 &   1.446 &         0.235 &        18.439 &           4.104 &         1.194 &         -0.774 &         -7.623 \\
 8.521 & 0.370 & 1.918 & -0.480 &  -5.214 &  12.863 &   2.266 &         0.820 &         7.746 &           2.560 &        -0.098 &         -0.890 &         -2.846 \\
 8.521 & 0.370 & 1.918 & -0.392 &  -2.918 &  14.490 &   0.311 &         0.430 &        14.241 &           3.698 &         0.826 &         -0.862 &         -5.817 \\
 8.521 & 0.610 & 2.366 & -0.764 &  -0.987 &  -0.767 &   1.427 &         0.846 &        22.534 &          13.381 &         4.285 &         -3.302 &        -17.062 \\
 8.521 & 0.612 & 2.371 & -0.904 &  -1.249 &   1.726 &  -0.867 &         0.193 &         3.512 &           3.172 &         0.742 &         -0.740 &         -3.106 \\
 8.521 & 0.615 & 2.375 & -1.048 &  -1.181 &   6.689 &  -3.413 &         0.161 &         2.180 &           2.470 &         0.459 &         -0.552 &         -2.003 \\
 8.521 & 0.616 & 2.379 & -1.373 &  -2.426 &   5.341 &   0.875 &         0.439 &         1.361 &           2.616 &         0.278 &         -0.737 &         -1.453 \\
 8.521 & 0.617 & 2.377 & -1.190 &  -2.099 &   3.914 &   1.694 &         0.245 &         1.912 &           2.948 &         0.401 &         -0.671 &         -2.004 \\
 8.847 & 0.482 & 2.309 & -0.385 &   4.663 &  16.563 & -12.772 &        24.800 &       197.164 &         133.490 &        69.492 &        -57.109 &       -160.615 \\
 8.847 & 0.483 & 2.311 & -0.446 &  -0.632 &   3.471 &   1.516 &         5.239 &        32.840 &          28.507 &        12.549 &        -11.947 &        -29.513 \\
 8.847 & 0.485 & 2.315 & -0.508 &  -3.060 &   1.339 &   4.197 &         3.836 &        17.411 &          18.690 &         7.323 &         -8.084 &        -16.785 \\
 8.847 & 0.486 & 2.317 & -0.570 &  -1.961 &   5.909 &   1.328 &         4.429 &        13.749 &          16.213 &         5.817 &         -7.862 &        -13.116 \\
 8.847 & 0.486 & 2.319 & -0.655 &  -7.329 &   2.650 &   6.605 &         4.030 &         6.017 &           9.520 &         1.643 &         -5.323 &         -5.131 \\
 8.851 & 0.497 & 2.121 & -0.410 &  -1.580 &  -3.137 &   1.537 &         0.718 &       349.240 &          51.819 &        15.153 &         -5.874 &       -132.602 \\
 8.851 & 0.501 & 2.128 & -0.480 &  -2.234 &   2.023 &   0.865 &         0.228 &        49.731 &          10.236 &         2.598 &         -1.262 &        -21.521 \\
 8.851 & 0.504 & 2.135 & -0.548 &  -2.984 &   8.501 &   1.573 &         0.259 &        26.770 &           6.979 &         1.731 &         -1.002 &        -12.708 \\
 8.851 & 0.506 & 2.138 & -0.614 &  -3.908 &   9.782 &   4.305 &         0.337 &        22.680 &           7.206 &         1.330 &         -0.973 &        -11.665 \\
 8.851 & 0.508 & 2.142 & -0.707 &  -3.673 &  18.717 &  -0.093 &         0.447 &        11.303 &           3.918 &         0.710 &         -0.740 &         -5.712 \\
10.591 & 0.369 & 2.128 & -0.209 &  -1.629 &  18.226 &  -0.828 &         1.988 &       317.897 &          36.895 &        24.257 &         -8.367 &       -106.116 \\
10.591 & 0.370 & 2.133 & -0.272 &  -2.133 &  29.686 &  -2.671 &         0.533 &        65.893 &          10.173 &         4.862 &         -2.112 &        -24.319 \\
10.591 & 0.371 & 2.138 & -0.483 &  -2.118 &  15.254 &  -1.618 &         1.191 &        20.746 &           4.516 &        -0.345 &         -1.264 &         -6.909 \\
10.591 & 0.372 & 2.138 & -0.336 &  -2.420 &  23.001 &  -3.923 &         0.393 &        38.624 &           6.906 &         2.251 &         -1.277 &        -14.776 \\
10.591 & 0.373 & 2.140 & -0.399 &  -2.882 &   6.871 &   0.556 &         0.621 &        34.084 &           6.696 &         1.344 &         -1.253 &        -12.933 \\
10.591 & 0.608 & 2.905 & -0.791 &  -2.117 &   0.404 &   1.683 &        15.910 &         3.673 &          37.487 &         7.533 &        -24.289 &        -11.525 \\
10.591 & 0.609 & 2.907 & -0.912 &   0.306 &   1.666 &  -1.912 &         2.970 &         0.446 &           6.897 &         0.993 &         -4.456 &         -1.530 \\
10.591 & 0.611 & 2.912 & -1.037 &  -0.089 &   2.729 &  -2.292 &         1.667 &         0.230 &           3.948 &         0.383 &         -2.491 &         -0.638 \\
10.591 & 0.613 & 2.915 & -1.163 &   0.478 &   3.517 &  -3.356 &         2.054 &         0.257 &           3.943 &         0.033 &         -2.674 &         -0.262 \\
10.591 & 0.613 & 2.917 & -1.328 &  -1.952 &   3.271 &  -0.367 &         2.374 &         0.484 &           3.158 &        -0.641 &         -2.511 &          0.392 \\
10.992 & 0.494 & 2.653 & -0.432 &   2.230 &   9.396 &  -3.736 &        14.674 &        50.633 &          57.667 &        26.937 &        -28.887 &        -53.245 \\
10.992 & 0.498 & 2.663 & -0.526 &   2.113 &   7.448 &  -4.848 &         4.001 &         9.605 &          14.664 &         5.644 &         -7.418 &        -11.202 \\
10.992 & 0.498 & 2.665 & -0.856 &  -8.611 &   8.704 &   5.592 &         5.203 &         4.212 &           5.014 &        -2.748 &         -4.394 &          0.682 \\
10.992 & 0.499 & 2.666 & -0.698 &  -0.050 &   6.829 &  -3.897 &         5.231 &         5.042 &           9.956 &         0.949 &         -6.359 &         -4.004 \\
10.992 & 0.499 & 2.668 & -0.613 &  -2.138 &   4.206 &   2.019 &         3.824 &         5.002 &          10.463 &         3.070 &         -5.913 &         -6.159 \\
\hline
\end{tabular}
    \caption{Table of kinematic bins, and CFF results for the ``difference" fit. The CFF results shown are the mean value as well as non-redundant covariances between the the CFFs. 
    Means and covariances are calculated using samples generated with Markov Chain Monte Carlo (MCMC) upon the difference likelihood, Eq.\eqref{eqn:likelihood_crosssectionDIFF}.
    }
\label{tab:results}
\end{table}

\clearpage
\subsection{Correlations study}
\label{subsec:covariance}
Our numerical analysis shows that, based on the set of data examined, we can extract quantitatively only three CFFs, $\Re e{\cal H}$, $\Re e{\cal E}$, and $\Re e \widetilde{{\cal H}}$. Furthermore, we always have covariant results obtained through both likelihood methods. 
Here we investigate possible causes for the covariance, and we suggest ways around it. 
Throughout this work, we singled out various sources that could generate the strong correlations seen in the analysis,  one by one showing their impact. We ruled out various possibilities while singling out which could be the most probable cause of correlation effects. Having ruled out aspects of the experimental data in the previous sections, the characteristic correlation structure that emerges from our analysis could be due to: 
\begin{enumerate}
    \item cross section formulation and its specific $t$ dependence
    \item perturbations due to non-leading terms/twist-three contributions to the cross section 
    \item asymmetry in size of the kinematic coefficients
\end{enumerate}

\subsubsection{Kinematic dependence of cross section}
\label{subsubsec:kinematic_dep}
One possible cause for covariant results is the form of the interference term of the cross section formula itself in the specific regime of interest. 
From Eq.\eqref{eq:sigmaINT1}, one can see that $\Re e {\cal H}$ and $\Re e {\cal E}$ cannot be simultaneously constrained, {\it i.e.} they are degenerate,  
in the regime where $F_1(t) \approx  \tau F_2(t)$. 
The ratio $F_1(t) \approx  \tau F_2(t)$ is shown in Figure \ref{fig:F1F2plot} where one can clearly see that this value is approached for at larger values of $-t$ than  the range of present experimental data (their values of $-t$ do not exceed 2 GeV$^2$). Therefore this source of correlation can be safely ruled out.  

\fixedfigure{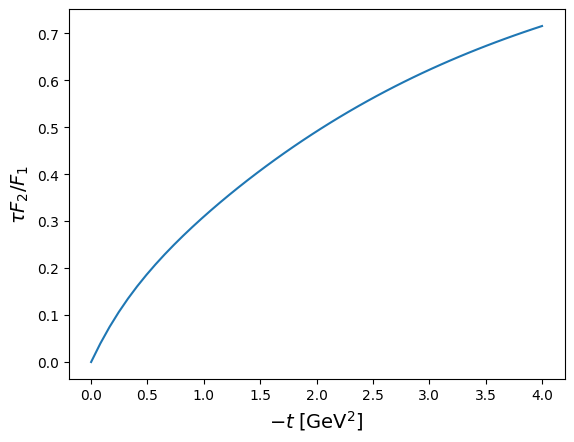}
{Ratio $\tau F_2/F_1$.  One can see that this quantity is not close to 1 for the present range of data ($-t < 2$ GeV$^2$) and thus it is unlikely to be the cause of the degeneracy of our CFF parameter extraction.}
{fig:F1F2plot}

\subsubsection{Beyond the leading order}
We investigated the role of non leading order corrections, specifically whether omitting twist-three terms in the QCD formulation leads to inconsistent results.
The reduced cross section given by,
\begin{equation}
    \sigma_\text{Reduced} \equiv \dfrac{Q^2\left(\sigma_{TOT} - \sigma_{BH}\right)}{\Gamma}
\end{equation}
is expected to be mildly $Q^2$ dependent in a QCD factorized framework, displaying logarithmic scaling violations at finite $Q^2$.
Nevertheless, it has been known for long that a tension exists between the leading -- logarithmic -- and subleading, $Q^2$ inverse powers dominated, twist contributions in QCD (we refer the reader to \cite{Pennington:1982kr,CTEQ:1993hwr} for earlier reviews on the subject, to \cite{Yang:1999xg} for an experimental study, and to \cite{QCDSFUKQCDCSSM:2022ncb} for a recent lattice QCD exploration). Higher twist effects are expected to be present in the large $x_{Bj}$ region where Jefferson Lab data are concentrated, therefore many analyses of DVES data have been concerned with their determination \cite{Defurne:2015kxq,Defurne:2017paw,Georges:2018kyi,JeffersonLabHallA:2022pnx}. 

\begin{figure}[h!]
    \includegraphics[width=0.6\linewidth]{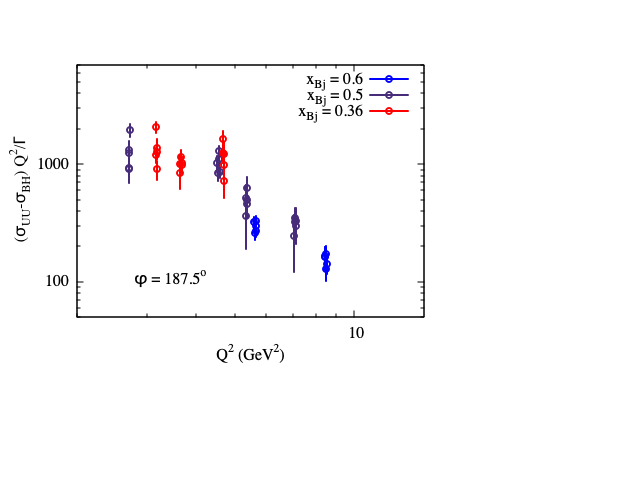}
    \caption{Example of reduced (dimensionless) cross section data from the Jefferson Lab Hall A collaboration \cite{Georges:2018kyi} as a function of $Q^2$ for different $x_{Bj}$ in the kinematic bin $\phi=187.5^{\rm o}$. The $x_{Bj}$ values are indicated in color. For each $x_{Bj}$ and $Q^2$ bin, several values of $t$ are available, yielding the spread of points in the figure. Similar trends are found for the other measured values of $\phi$. }
\label{fig:ReducedCrossSectionQsqDependence}
\end{figure}

In Fig.~\ref{fig:ReducedCrossSectionQsqDependence}, we plot the reduced cross section using data reported by the Jefferson Lab Hall A collaboration \cite{Georges:2018kyi} with respect to $Q^2$. 
Each $x_{Bj}$ bin value contains data at different values $E$ and $t$, for a given $\phi$ value, although a similar behavior is shown for the other kinematic bins.
The spread in the reduced cross section values for each chosen kinematics is due to the $t$ dependence of the data. From the trend of the data in $Q^2$, we deduce that the effect of $Q^2$ dependence other than logarithmic, such as {\it e.g.} the one given by higher twist terms cannot be ruled out. 

By performing a likelihood analysis along the lines of the leading order one, we are faced with the situation where for the next non leading order contribution -- which is expected to be of twist three \cite{Kriesten:2019jep} --  there are eight new complex CFFs, from the eight twist-three GPDs, which translate, in general, into sixteen unknowns \cite{Kriesten:2019jep,Kriesten:2020wcx,Meissner:2009ww}. 

In this case, the DVCS and DVCS-BH interference contributions to the cross section read,
\begin{eqnarray}
\label{eq:sigmaDVCS_tw3}
\sigma_{DVCS}  & = & \frac{\Gamma}{Q^2(1-\epsilon)} \left\{ F_{DVCS}^{tw 2} +  \sqrt{\frac{t_0-t}{Q^2}} \sqrt{\epsilon(1+\epsilon)} \cos \phi  \, F_{DVCS}^{tw 3} \right\}
\\
\label{eq:Int_FUU3}
\sigma_{\cal I}  & = &    \frac{\Gamma}{Q^2} \left\{ F_{\cal I}^{tw 2} +  \sqrt{\frac{t_0-t}{Q^2}}  F_{\cal I}^{tw 3} \right\} 
\end{eqnarray}
where $F_{DVCS}^{tw 2}$ and $F_{\cal I}^{tw 2}$, the unpolarized electron-proton twist two contributions,  can be read off Eqs.\eqref{eq:sigmaDVCS1} and \eqref{eq:sigmaINT1}.

The unpolarized electron-proton twist three contributions read,
\begin{eqnarray}
F_{DVCS}^{tw 3} & = & - 2 \,  (1-\xi^2) \, 
\Re{\rm e} \Bigg\{  \left(2 \widetilde{\cal H}_{2T} + {\cal E}_{2T} + 2 \widetilde{\cal H}'_{2T} + {\cal E}'_{2T} \right)^{*}\Big( {\cal H}-\frac{\xi^2}{1-\xi^2} \mathcal{E} \Big)  \nonumber \\
&+& \Big(\mathcal{H}_{2T} + \tau \widetilde{\mathcal{H}}_{2T} + \mathcal{H}_{2T}' +  \tau \widetilde{\mathcal{H}}_{2T}'  \Big)^{*}\Big( \mathcal{E} - \xi \widetilde{\mathcal{E}} \Big)
\nonumber \\
&- & 2 \xi \,\left(\widetilde{\cal E}_{2T} + \widetilde{\cal E}'_{2T}\right)^{*} \Big(\widetilde{\cal H} - \frac{\xi^2}{1-\xi^2} \widetilde{\mathcal{E}}\Big) 
+ \frac{\xi}{1-\xi^2} \Big(\widetilde{\mathcal{E}}_{2T} - \xi \mathcal{E}_{2T} + \widetilde{\mathcal{E}}_{2T}' - \xi \mathcal{E}_{2T}' \Big)^{*} \Big( \mathcal{E} - \xi \widetilde{\mathcal{E}} \Big)
\nonumber \\
&+&  \frac{\tau}{4}   \Big(\mathcal{\widetilde{H}}_{2T} + \mathcal{\widetilde{H}}_{2T}'\Big)^{*} \Big( \mathcal{E} + \xi \widetilde{\mathcal{E}}\Big)   \Bigg\}
\end{eqnarray}

\begin{eqnarray}
F_{\cal I}^{tw 3} & = & A^{(3)}_{\cal I} \Re e \Bigg\{ F_{1} \Big(  (2   \mathcal{\widetilde{H}}_{2T} +   \mathcal{E}_{2T}) - (2  \widetilde{\mathcal{H}}'_{2T} +  \mathcal{E}_{2T}') \Big)
 +  F_{2} \Big(  (  \mathcal{H}_{2T} + \tau  \mathcal{\widetilde{H}}_{2T}) - ( \mathcal{H}_{2T}' + \tau  \widetilde{\mathcal{H}}_{2T}' ) \Big) \Bigg\} 
 \nonumber \\
 &+ &  B^{(3)}_{\cal I} (F_1+F_2) \, \Re e  \Big\{  \widetilde{\mathcal{E}}_{2T} - \xi {\cal E}_{2T}  - \xi (\widetilde{\mathcal{E}}_{2T}' - \xi {\cal E}'_{2T}) + \xi {\cal E}_{2T} - \xi^2 E'_{2T} \Big\} \nonumber 
 \\
 &+ &  C^{(3)}_{\cal I}  (F_1+F_2)  \,   \Re e \Bigg\{ 2\xi   ( \mathcal{H}_{2T} -  \mathcal{H}_{2T}')- \tau \Big(  \widetilde{\mathcal{E}}_{2T}  - \xi \mathcal{E}_{2T}  - ( \, \widetilde{\mathcal{E}}_{2T}'    
       + \xi \mathcal{E}_{2T}'  ) \Big) \Bigg\}
\end{eqnarray}
In the unpolarized cross section we single out only three vector coupling GPDs, {\it i.e.} only an additional one at twist three, corresponding to the quark-proton polarizations, $UU, UT, UL$ described respectively, by the combinations, 
\begin{itemize}
    \item $2\widetilde{H}_{2T} + E_{2T}$, the twist three polarization configuration analogous to the twist two $H$, unpolarized quark in unpolarized proton, $UU$;
    \item $ H_{2T} + \, \tau \widetilde{H}_{2T} $, twist three polarization configuration analogous to the twist two $E$, unpolarized quark in transversely polarized proton, $UT$;
    \item $\widetilde{E}_{2T} - \xi E_{2T}$, corresponding to an unpolarized quark in longitudinally polarized proton, $UL$, that has no analogous at twist two due to parity conservation (see discussion in Refs.\cite{Courtoy:2013oaa,Rajan:2016tlg,Raja:2017xlo}). This term is important to the overall picture because it describes the orbital component of angular momentum \cite{Kiptily:2002nx,Hatta:2011ku,Raja:2017xlo}. 
\end{itemize}
Similarly, we have three terms with axial-vector couplings described by the GPDs and CFFs with a $\prime$ superscript,
\begin{itemize}
    \item $2\widetilde{H}'_{2T} + E'_{2T}$, the twist three polarization configuration analogous to $\widetilde{H}$, unpolarized quark in unpolarized proton, $LL$;
    \item $ H'_{2T} + \, \tau \widetilde{H}'_{2T} $, twist three polarization configuration analogous to $E$, unpolarized quark in transversely polarized proton, $LT$;
    \item $\widetilde{E}'_{2T} - \xi E'_{2T}$, corresponding to a longitudinally quark in an unpolarized  proton, $LU$, that also has no analogous at twist two due to parity conservation (see discussion in Refs.\cite{Courtoy:2013oaa,Rajan:2016tlg,Raja:2017xlo}). This term describes the quark spin orbit contribution \cite{Kiptily:2002nx,Hatta:2011ku,Raja:2017xlo}. 
\end{itemize}

(see \cite{Kriesten:2020wcx}, for a more extensive discussion of the physical meaning of these terms). 

\begin{figure}[H]
    \centering
    \includegraphics[width=0.6\linewidth]{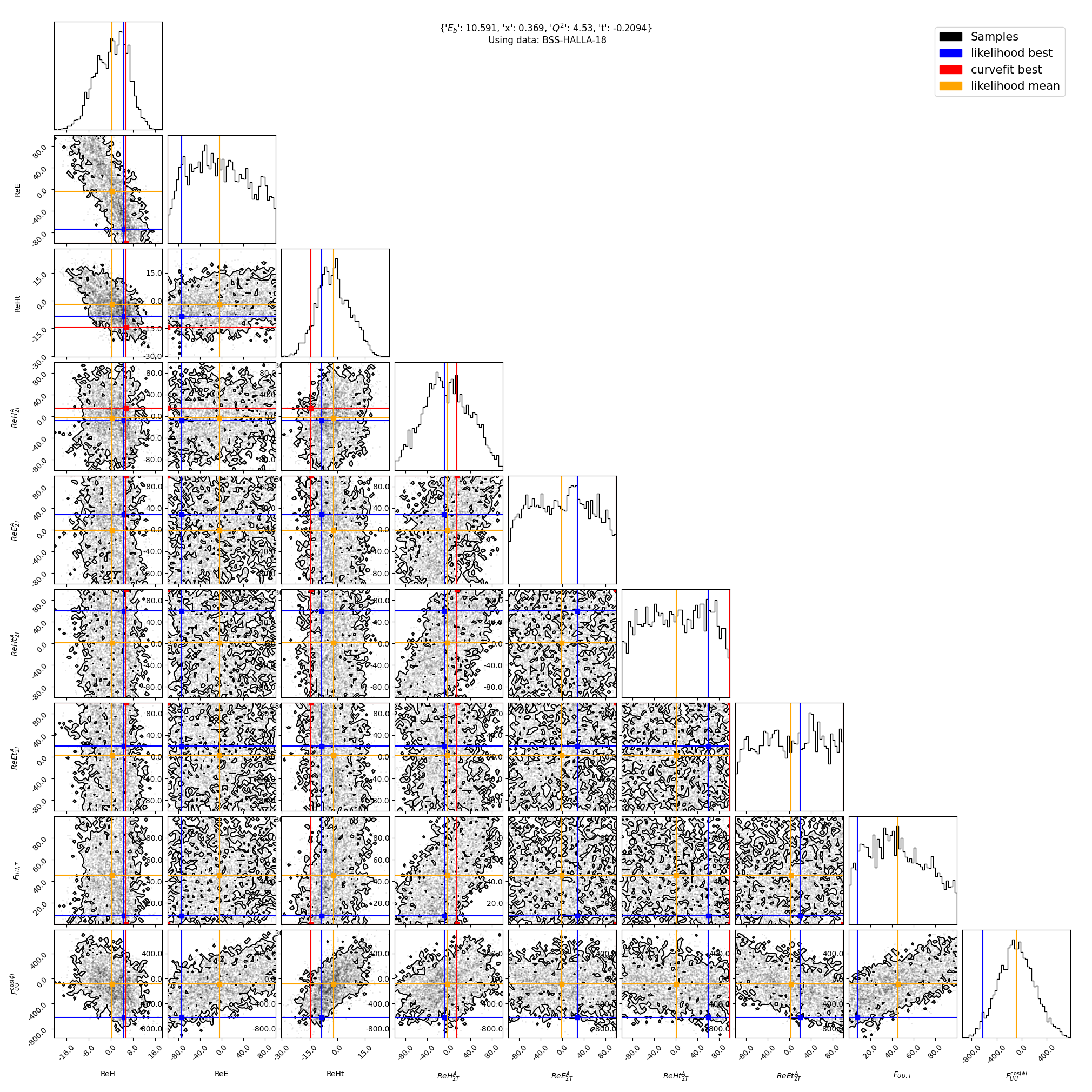}
    \includegraphics[width=0.3\linewidth]{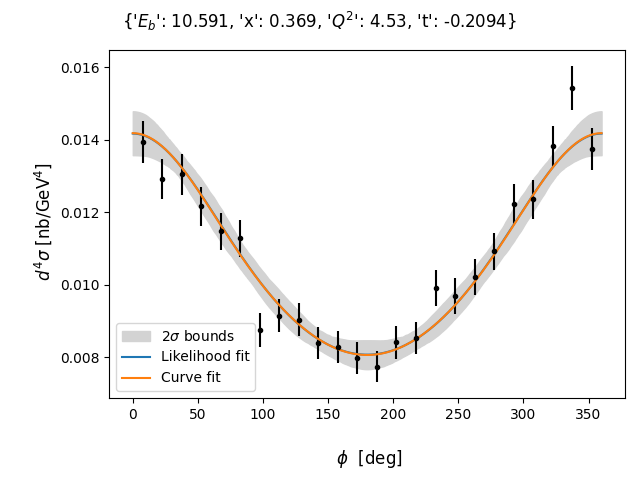}
    \caption{Left: Corner plot with twist-3 CFFs included. CFF parameters bounded by hyperbox $[-100, 100]$ while $F_{UU,T}$ and $F^{\cos(\phi)}_{UU}$ bounded between $[-10000, 10000]$.
    5000 samples are included in this graph.
    The best fit likelihood occurs at: 
    [   4.57,   -73.52,   -8.40,   -8.42,   28.13,   60.34,   20.04,    8.17,  -623.13].
Right: Likelihood fit with twist-3 CFFs included, corresponding to the {\it lhs} panel. 
    Error bands are generated by drawing 1000 curves on this plot and finding the cross section values, for each $\phi$ value, located above/below 95\% of the curves.
    The orange curve corresponds to the best fit CFF parameters which draw a curve through the data.
    The blue curve (which appears superimposed to the orange one) corresponds to the MCMC sample of CFF parameters which has the largest likelihood.
    }
    \label{fig:corner_UU_TW3_v1}
\end{figure}

The coefficients, $A^{(3)}_{\cal I}$, $B^{(3)}_{\cal I}$, $C^{(3)}_{\cal I}$  are written in terms of four-vector products involving all relevant kinematics variables. Similarly to the twist-two case, they can be expressed in terms of the set of variables ($Q^2$, $x_{Bj}$, $t$, $\phi$). Their specific expressions are rather lengthy and are given explicitly in Ref.\cite{Kriesten:2019jep}. 
For the present analysis it is important to point out that the twist-three DVCS cross section, Eq.\eqref{eq:sigmaDVCS_tw3}, depends on $\cos \phi$, thus complicating the outcome of a possible analysis in the canonical framework. Results are plotted in Figure \ref{fig:corner_UU_TW3_v1}. 


\textcolor{blue}{
We have derived the canonical likelihood composed of twist-3 cross section and unpolarized DVCS data. We attempted to constrain the associated twist-3 CFFs using MCMC.
The results of applying such an MCMC are shown in Figure \ref{fig:corner_UU_TW3_v1}.
One can clearly see the resulting twist-3 CFF parameters are degenerate.
The resulting MCMC samples are entirely dependent upon a choice of prior. 
Thus the twist-3 CFFs are not fully bounded using a canonical likelihood analysis (which we performed). 
We relegate any approaches to tackle the twist-3 degeneracy problem to future work.
}

\textcolor{blue}{\subsection{Cross section modeling}
\label{subsec:BKM}
Finally, the expressions for the DVCS and BH cross sections available in the literature differ from one another, and have been notoriously difficult to compare analytically. We single out three major approaches where the calculation of the process, $e p \rightarrow e' p' \gamma'$, was fully completed, namely Refs.\cite{Belitsky:2001ns,Belitsky:2010jw} (BKM), Refs.\cite{Braun:2012hq,Braun:2014sta} (Braun {\it et al.}), and Refs.\cite{Kriesten:2019jep,Kriesten:2020wcx,Kriesten:2020apm} (UVA). 
We performed a direct comparison of the formalism of BKM and UVA, since these collaborations published complete formulations of their results. 
While the BH, Eq.\eqref{eq:sigmaBH1}, and DVCS, Eq.\eqref{eq:sigmaDVCS1}, cross section contributions agree between the two groups, the interference term  exhibits  differences in the  $\phi$ dependence at fixed kinematics, $(E,t,x_{Bj},Q^2)$. These differences were also highlighted in Ref.\cite{Kriesten:2020wcx}.
Here, we repeated the likelihood analysis at the same kinematic bin showed in Fig.\ref{fig:3CFF}, by working out the kinematic coefficients, $A_{\cal I}, B_{\cal I}, C_{\cal I}$ defined in  Eq.\eqref{eq:sigmaINT1}, starting from the BKM expressions. 
}

\textcolor{blue}{The figure clearly shows that both formalisms can be used to fit the cross section: performing a curve fit, we found that BKM can fit the unpolarized cross section, equivalently to the UVA approach (Figure \ref{fig:BKMComp_corner}, Right). However, the main result of this analysis, is the clear degeneracy displayed in the BKM results (Fig.\ref{fig:BKMComp_corner}, left): in other words, our analysis clearly indicates that using the BKM formalism does not allow one to extract CFFs from DVCS unpolarized data.
\begin{figure}[h]
    \includegraphics[width=9cm]{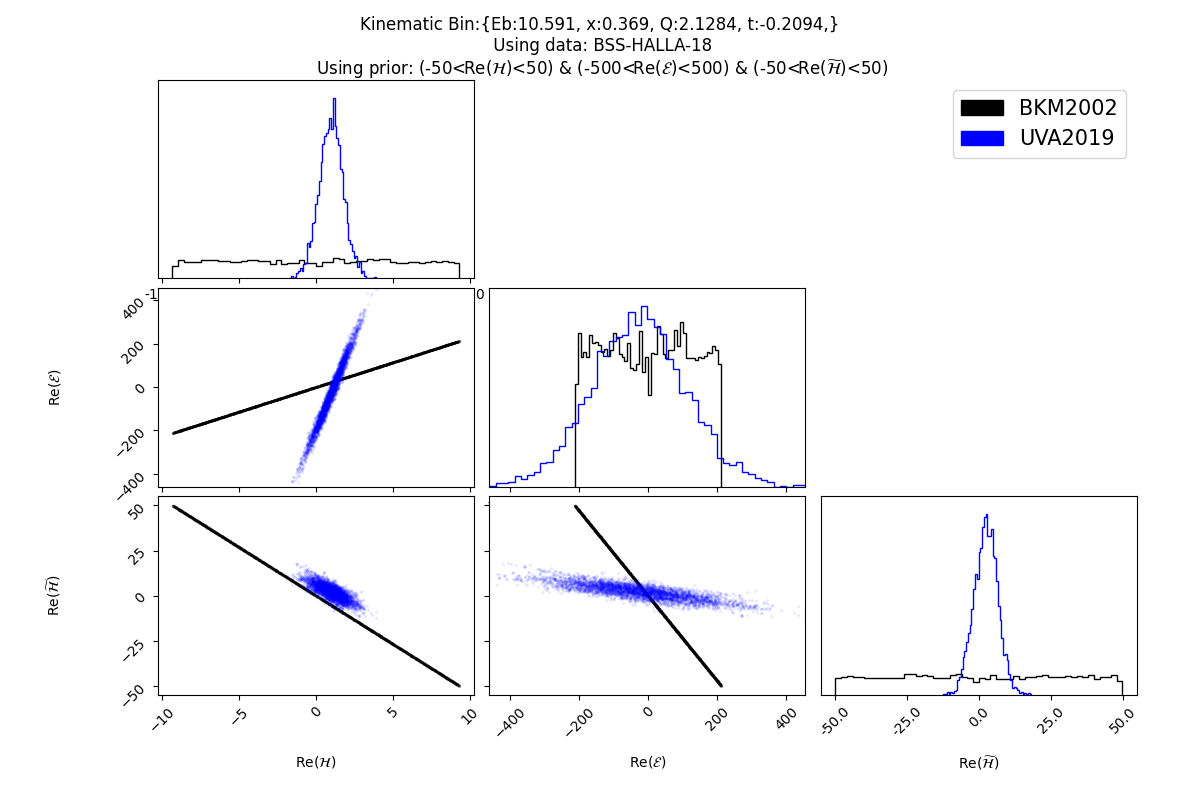}
    \hspace{1cm}
    \includegraphics[width=6cm]{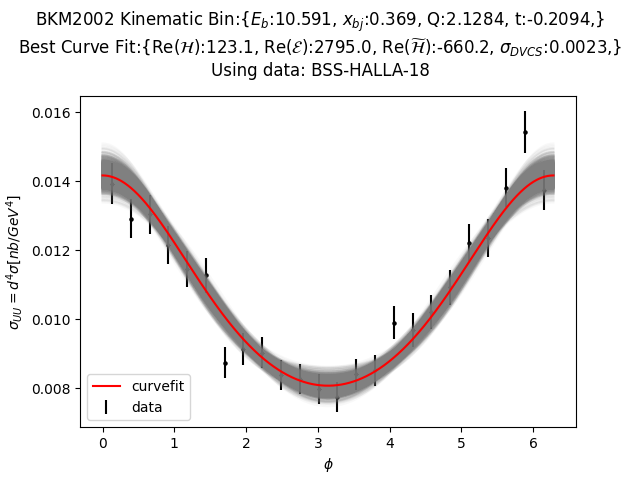}
    \caption{Left:
    Corner plot comparison of 3 CFFs, $\Re e {\cal H}$, $\Re e {\cal E}$, $\Re e \widetilde{\cal{H}}$, calculated in the UVA (blue) and BKM (black) formalisms in the kinematic bin: $E=10.59$ GeV, $t=-0.21$ GeV, $x_{Bj} = 0.37$, $Q^2=4.5$ GeV$^2$. The UVA analysis corresponds to the values displayed in Fig.\ref{fig:3CFF}. The BKM results clearly show degeneracy (see text). Right: results from using the BKM formalism only. The spaghetti plot selects 1000 MCMC samples and then draws 1000 corresponding curves. Separately a least-squares curve-fit is found and shown in red.
}
    \label{fig:BKMComp_corner}
\end{figure}
%
%
}

\textcolor{blue}{In Figure \ref{fig:BKM_Comp_ABC} we hint at a possible reason for the degeneracy of the BKM results, in that the expressions of $A_{\cal I}, B_{\cal I}$ and $C_{\cal I}$, that would allow one to separate the different CFF results, seem to be dominated by the {\it same} $\phi$ modulation for BKM, whereas the diverse behavior displayed in the UVA formulation allows one to separate the terms in the analysis. The appearance of a same modulation for BKM might be a natural consequence of using the harmonic expansion that said formalism is based on. The dependence on the phase $\phi$, emerging from the the UVA analysis of in $2 \rightarrow 3$ scattering events was discussed in detail in {\it e.g.} Ref.\cite{Kriesten:2020wcx}.
%
\begin{figure}[h]
    \centering
    \includegraphics[width=0.33\linewidth]{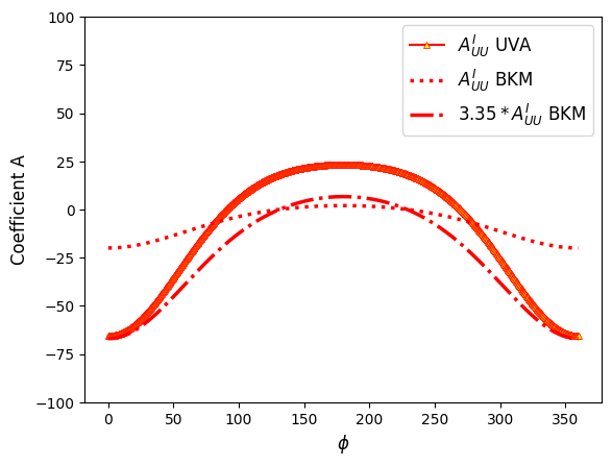}
    \includegraphics[width=0.33\linewidth]{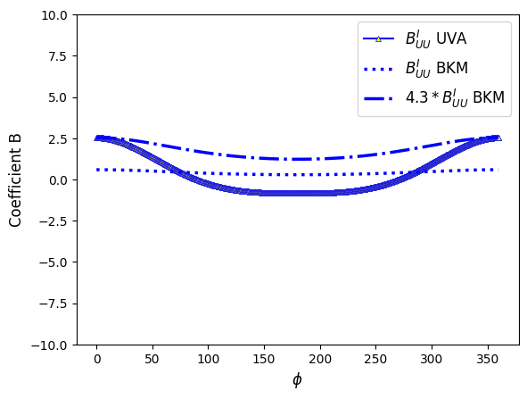}
    \includegraphics[width=0.33\linewidth]{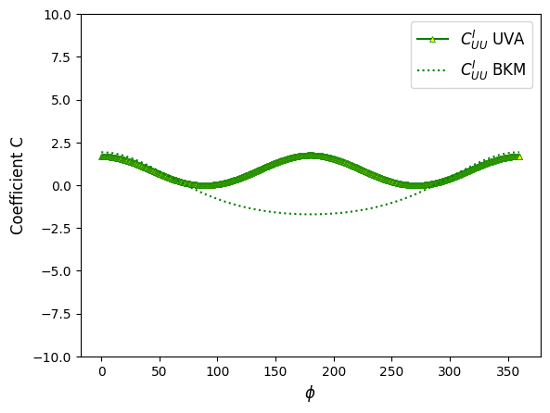}
    \caption{Kinematic coefficients of the DVCS-BH interference term , $A$, $B$, $C$, Eq.\eqref{eq:sigmaINT1}, evaluated at: $E_{beam}=10.591$ (GeV), $x_{Bj}=0.369$, $Q^2=4.53$ $({\rm GeV}^2)$, $t=-0.2094$ $({\rm GeV}^2)$ under UVA and BKM formalism.}
    \label{fig:BKM_Comp_ABC}
\end{figure}
}

\section{Conclusions and Outlook}
\label{sec:conclusions}
We presented a new paradigm for the analysis of deeply virtual exclusive photoproduction off a proton target, $ep \rightarrow e' p' \gamma'$, using a model for likelihood-based inference in high dimensional data settings. 

The deeply virtual Compton scattering contribution to the cross section, where the photon is produced at the proton vertex, is described in terms of Compton form factors which, in a QCD factorized scenario can be written as convolutions of GPDs, which are believed to contain important information on the proton 3D structure. The extraction of Compton form factors from data represents, therefore, the first step in obtaining the latter from data.

We obtained joint (statistically) covariant results for Compton form factors using a twist-two cross section model
for the unpolarized process for which the largest amount of data with all the kinematic dependences are available from 
corresponding datasets from Jefferson Lab.
%
Our analysis provides a method which derives a joint likelihood 
of the latter, 
for each observed combination of the kinematic variables defining the reaction,
$[E, x_{B}, t,  Q^2]$, and varying azimuthal angles $\phi$.
Based on the observation that the unpolarized twist-two cross section likelihood fully constrains only three of the Compton form factors,
the derived difference-model likelihoods were explored using Markov chain Monte Carlo (MCMC) methods.
Error bars, covariances and associated joint $n-\sigma$ confidence contours were evaluated.
Our main results for the twist-two cross section are presented in a 45 rows table (Table \ref{tab:results}), one row for each unique kinematic bin from our available unpolarized DVCS dataset \cite{GepardGithub} \cite{Georges:2018kyi}. Results are also shown using corner plots to highlight the impact of the covariances in the analysis. 

What can be done to reduce the size of the $n-\sigma$ contours? 
Using a more ``well behaved" set of data does not substantially help. 
At the same time we see that higher twists are important. 
This casts doubts on the validity of a QCD factorized description approach \cite{Collins:1998be} for the present kinematic setting, and motivates both further investigations, {\it e.g.} at EIC kinematics which guarantees a large $Q^2$ span. Previous HERA data \cite{H1:2001nez} as precursors to the EIC, could also provide additional information, as it would extending the analysis to include constraints from other types of data (considering  other polarization configurations, {\it e.g.} longitudinally and transversely polarized targets and recoil target polarization) for which at present, only simulations are made available. 

One error reducing method we plan to explore is known as ``outlier pruning" \cite{Hogg:2010yz}. 
The procedure involves designing an analysis that allows for each data point considered to have an effective weighting based on the probability of the point being an outlier. 
Then these outlier probabilities are marginalized in the final likelihood analysis in the MCMC step. 
The method either introduces many additional parameters which must be marginalized over, or requires clever math methods which roll the additional parameters into few parameters. 
We plan to attempt a statistically sound outlier pruning in our next paper.

The impact of the twist-three corrections to the analysis was explored but the current status of the analysis is incomplete: while we included the calculations of the twist-three corrections which introduce sixteen additional Compton form factors, and calculated an introductory twist-three canonical likelihood, 
we note that a proper likelihood analysis treatment in this case requires careful assessment of degeneracy in the Compton form factors. Thus we relegate a careful twist-three parameter degeneracy investigation to future work. 

\textcolor{blue}{Finally, we took into account different approaches to writing the DVCS cross section, namely, BKM \cite{Belitsky:2001ns,Belitsky:2010jw}, Braun {\it et al.} \cite{Braun:2012hq,Braun:2014sta} and UVA \cite{Kriesten:2019jep,Kriesten:2020wcx}. We found that BKM's formalism results in a much larger covariance  than the UVA case, leading to degeneracy. This finding emphasizes the importance of carefully considering the phase $\phi$ dependence of deeply virtual exclusive processes, including DVCS, where at least three particles are measured in the final state.}
Once the basic physics framework \textcolor{blue}{including QCD factorization} is established, we will be in a position to constrain the CFFs using other sets of data to including parton distributions from deep inelastic inclusive scattering and opening to lattice QCD.


In summary, our present analysis lays out the statistics foundation of our new framework which leverages state of the art ML algorithms to carry out efficiently Bayesian inference in high-dimensional settings. Companion manuscripts to the present one, addressing the connection with ML can be found in Refs.\cite{Almaeen:2024guo,Hossen:2024qwo}.   

Previous analyses \cite{JeffersonLabHallA:2022pnx} do not agree with our results. The analyses of deeply virtual exclusive experiments conducted so far do not provide a publicly available, common set of benchmarks of their statistical approaches, and thus we are not in a position to attempt to reproduce existing results for comparison. Nevertheless, Fig.\ref{fig:Comp_Georges} shows a comparison of our analysis with published results. 
From our analysis we conclude that in order to validate a QCD factorized scenario from which the observables for the 3D structure of the proton could be extracted,
it is important to perform measurements for different target polarizations, both longitudinal and transverse, as well as at the lowest $t/Q^2$ ratio and widest $Q^2$ range possible.


\acknowledgments 
This work was completed grant by the EXCLAIM collaboration under the Department of Energy grant DE-SC0024644. 
We also thank Stefan Baessler for discussions.   

\appendix
\section{BH coefficients}
\label{app:A}
The BH coefficients in Eq.(\ref{eq:sigmaBH1}) read,
\begin{align}
A_{BH}   = &\frac{{8}\,M^2}{t \myprod{k}{q'} \myprod{k'}{q'}}
\Bigg[  
4 \tau 
 \Big(
 \myprod{k}{P}^2
 +
 \myprod{k'}{P}^2
 \Big)  
 -(\tau +1)
\Big(
 \myprod{k}{\Delta}^2
 +
 \myprod{k'}{\Delta}^2
\Big)
\Bigg] 
\label{eq:Aunpol}\\
B_{BH} = & 
\frac{{16} \,M^2}{t \myprod{k}{q'} \myprod{k'}{q'}}
\Big[
 \myprod{k}{\Delta}^2
 +
 \myprod{k'}{\Delta}^2
\Big]  \,,
\label{eq:Bunpol}
\end{align}
where  $P=(p+p')/2$,  $\tau=-t/4M^2$. The four vector products, $(ab)=a_o b_o - {\bf a} \cdot {\bf b}$, can be evaluated by for either the fixed target or collider settings.

\section{Interference coefficients}
\label{app:B}
\noindent The BH-DVCS interference cross section coefficients in Eq.(\ref{eq:sigmaINT1}) read ($UU$ refers to the unpolarized electron beam scattering off the unpolarized target),
\begin{eqnarray}
  A_{UU}^{\cal I}  & = & -8 D_{+} \Big[ (k'P) \Big(2 k_T^2 -k_T\cdot q_T' - 2(kq')\Big)  
 +  (kP) \Big( 2 k'_T \cdot k_{T} + k'_T \cdot q'_{T} +2(k'q')\Big) - (k_T \cdot P_T)  \Big(2(kk') + (k'q') \Big) \nonumber \\
 & + &  (k_T' \cdot P_T) (kq') \Big] \cos \phi 
 - 8D_{-} \Big[ (Pq') \Big(2 k_T \cdot k'_{T}  + 2(kk')\Big) 
  - (k_T \cdot P_T) (k'q') - (k_T' \cdot P_T)(kq')\nonumber \\
  &+& (P_T \cdot q_T') (kk')  \Big] \cos \phi  \\
 B_{UU}^{\cal I} & = &   -4 \xi D_{+} \Big[ (k'\Delta) \Big(2 k_T^2 -k_T\cdot q_T' - 2(kq')\Big)  
 +  (k\Delta) \Big( 2 k'_T \cdot k_{T} + k'_T \cdot q'_{T} + 2(k'q')\Big) - (k_T \cdot \Delta_T)  \Big(2(kk') + (k'q') \Big) \nonumber \\
 & + &  (k_T' \cdot \Delta_T) (kq') \Big] \cos \phi 
 - 4\xi D_{-} \Big[ (\Delta q') \Big(2 k_T \cdot k'_{T}  + 2(kk')\Big) 
  - (k_T \cdot \Delta_T) (k'q') - (k_T' \cdot \Delta_T)(kq')\nonumber \\
  &+& (q_T' \cdot \Delta_T) (kk')  \Big] \cos \phi  \\
 C_{UU}^{\cal I} & = & 
 4\frac{D_{+}}{P^{+}}\Big[ -(k_{T}\cdot P_{T})(q'_{T} \cdot \Delta_{T})k^{\prime +} + (k_{T} \cdot \Delta_{T})(P_{T}\cdot q'_{T})k^{\prime +} + (k'_{T}\cdot P_{T})(q'_{T} \cdot \Delta_{T})k^{+} \nonumber \\
 &-&(k'_{T} \cdot \Delta_{T})(P_{T} \cdot q'_{T})k^{+} - 2(kk')(k'_{T} \cdot P_{T})\Delta^{+} -(kq')(k'_{T} \cdot P_{T})\Delta^{+} \nonumber \\
 &+& (k'q')(k_{T} \cdot P_{T})\Delta^{+} + 2(kk')(k'_{T} \cdot \Delta_{T})P^{+} + (kq')(k'_{T} \cdot \Delta_{T})P^{+} - (k'q')(k_{T} \cdot \Delta_{T})P^{+}\Big] \cos \phi  \nonumber \\
&+& 4 \frac{D_{-}}{t P^{+}}\Big[(k'q')(k_{T} \cdot P_{T})\Delta^{+} - (kk')(q'_{T} \cdot P_{T})\Delta^{+} + (kq')(k'_{T} \cdot P_{T})\Delta^{+} \nonumber \\
&-& (k'q')(k_{T} \cdot \Delta_{T})P^{+} + (kk')(q'_{T} \cdot \Delta_{T})P^{+} - (kq')(k'_{T} \cdot \Delta_{T})P^{+}\Big] \cos \phi 
\end{eqnarray}
where the ``$T$" subscript refers to the transverse component of the given vectors with respect to the $ep$ direction; the $+$ superscript in the vectors, $v^+=(P^+, \Delta^+, k^+, k'^+)$, refers to the combination: $v^+ = (v^o+v^3)/\sqrt{2}$; $D_+$ and $D_-$ are defined as,
\begin{equation} 
D_+ = \frac{(kq')-{(k'q'})}{2(k'q')(kq')}, \;\;\;\; \quad\quad\quad D_- = -\frac{(kq')+{(k'q')}}{2(k'q')(kq')}
\end{equation}
The coefficients also receive a contribution from the electromagnetic gauge invariance preserving terms, which only mildly modifies the values of the coefficients. A complete definition of the coefficients up to twist three is given in \cite{Kriesten:2020wcx}.

\bibliography{references}

\end{document}